\documentclass[superscriptaddress,reprint,amsmath,amssymb,aps,prb,floatfix]{revtex4-1}

\usepackage{xcolor} 
\usepackage{graphicx}
\usepackage{dcolumn}
\usepackage{upgreek}
\usepackage{bm}
\usepackage{hyperref}
\usepackage{floatrow}
\usepackage{times}
\usepackage[mathlines]{lineno}
\usepackage{comment}
\usepackage{physics}
\usepackage{amsmath}
\usepackage{subfiles}
\usepackage{multirow,tabularx}

\usepackage{etoolbox}
\makeatletter
\newcommand*{\newbibstartnumber}[1]{%
  \apptocmd{\thebibliography}{%
    \global\c@NAT@ctr #1\relax
    \addtocounter{NAT@ctr}{-1}%
  }{}{}%
}
\makeatother

\begin{document}

\title{
Dispersive Non-reciprocity between a Qubit and a Cavity 
}
\author{Ying-Ying Wang}
\affiliation{Department of Physics, University of Massachusetts-Amherst, Amherst, MA, USA}
\author{Yu-Xin Wang}
\affiliation{Pritzker School of Molecular Engineering, University of Chicago, Chicago, IL, USA}
\author{Sean van Geldern}
\affiliation{Department of Physics, University of Massachusetts-Amherst, Amherst, MA, USA}
\author{Thomas Connolly}
\email{Present address: Department of Applied Physics, Yale University, New Haven, CT, USA}
\affiliation{Department of Physics, University of Massachusetts-Amherst, Amherst, MA, USA}
\author{Aashish A. Clerk}
\affiliation{Pritzker School of Molecular Engineering, University of Chicago, Chicago, IL, USA}
\author{Chen Wang}
\email{wangc@umass.edu}
\affiliation{Department of Physics, University of Massachusetts-Amherst, Amherst, MA, USA}

\date{\today}
\begin{abstract}

The dispersive interaction between a qubit and a cavity is ubiquitous in circuit and cavity quantum electrodynamics.  It describes the frequency shift of one quantum mode in response to excitations in the other, and
in closed systems is necessarily bidirectional, i.e.~reciprocal.  Here, we present an experimental study of a non-reciprocal dispersive-type interaction between a transmon qubit and a superconducting cavity, arising from a common coupling to dissipative intermediary modes with broken time reversal symmetry.
We characterize the qubit-cavity dynamics, including asymmetric frequency pulls and photon shot-noise dephasing, under varying degrees of non-reciprocity by tuning the magnetic field bias of a ferrite component in situ.
Furthermore, we show that the qubit-cavity dynamics is well-described in a wide parameter regime by a simple non-reciprocal master-equation model, which provides a compact description of the non-reciprocal interaction without requiring a full understanding of the complex dynamics of the intermediary system.  Our result provides an example of quantum non-reciprocal phenomena beyond the typical paradigms of non-Hermitian Hamiltonians and cascaded systems.  

\end{abstract}
\maketitle

\section{Introduction}

Theoretical and experimental studies of non-reciprocity
are of great interest, both due to their fundamental implications for realizing exotic phases of matter~\cite{Fruchart_2021,Shankar_2022,Bowick_2022}, as well as their relevance to applications in classical and quantum information processing.  The most widely investigated non-reciprocal phenomena concerns the scattering matrix $S$ of input and output signals for a multi-port network, where the transmission coefficient is not invariant under the exchange of the source and the receiver, i.e.~$S_{ij}\neq S_{ji}$, with isolators and circulators being the canonical examples.  Realization of \textit{non-reciprocal scattering} in optical~\cite{Ruesink_2016,Fang_2017}, acoustic~\cite{Xu_2019,Nassar_2020}, and microwave domains~\cite{Sliwa_2015,Kamal_2011} with new techniques has been an intensive area of study.  On the other hand, the concept of non-reciprocity goes beyond the scattering properties of propagating linear modes, especially in quantum contexts where one often considers non-reciprocal interactions between stationary quantum subsystems.  When the stationary modes are linear and in the classical correspondence limit, their non-reciprocal interactions can be conveniently described by a non-Hermitian Hamiltonian~\cite{Fang_2017, Xu_2019,Helbig_2020,Ghatak_2020}. Such non-Hermitian dynamics have been shown in recent works to generate a plethora of new physical phenomena, including the non-Hermitian skin effect~\cite{Yao_2018,Helbig_2020,Ghatak_2020} and novel critical phenomena under monitored dynamics~\cite{Gopalakrishnan_2021,Kawabata_2023}.

Non-reciprocal interactions that are more uniquely quantum arise when the subsystems of interest include strongly nonlinear modes, which necessitates the use of master equations to describe the system dynamics. A well studied example is a cascaded network of resonant qubits~\cite{Stannigel_2012}, where the interaction between neighboring qubits is mediated by emission and absorption via 
a directional waveguide. The resuling effective interaction can be described as non-reciprocal transfer of single excitations. This cascaded model has been investigated theoretically, where the non-reciprocity leads to unique dynamics such as steady-state entanglement and dimerized many-body states.  Experimental development in chiral quantum optical platforms~\cite{Lodahl_2015,Sayrin_2015,Lodahl_2017} and waveguide circuit QED~\cite{Nicolas_2020,Kannan_2023,Guimond_2020,Joshi_2023} are expected to realize such resonant non-reciprocal phenomena in the near future.

It is natural to consider the generic properties of non-reciprocal interactions that go beyond simple one-way excitation transfer, as is expected if the relevant subsystems are non-resonant.  Are there generic theoretical models and experimental signatures for such interactions? This is particularly relevant in quantum device engineering where it is commonplace to utilize weak hybridization of disparate linear and nonlinear modes.  For example, the dispersive Hamiltonian between a qubit and a cavity, $\frac{\hbar\chi}{2} \hat{a}^\dagger\hat{a}\hat{\sigma}_z$, forms the cornerstone of circuit QED and superconducting quantum computation~\cite{Blais_2004}. Here the qubit or the cavity experiences a frequency shift in response to an excitation in one other, by the same amount $\chi$, exemplifying the reciprocal nature of the dispersive interaction in a closed quantum system.

Recently, a new class of non-reciprocal interaction in this category and distinct from cascaded quantum systems has been theoretically investigated~\cite{Yuxin_2023}; these can naturally arise from 
dispersive-type couplings in open quantum systems.  The simplest example is described by a single Lindblad dissipator 
$\mathcal{D}[\hat{a} e^{i\theta \hat{\sigma}_z}]$, 
which describes a one-way influence of a cavity mode $\hat{a}$ on a qubit $\hat{\sigma}_z$ in terms of a phase or frequency shift. Given the ubiquity of standard dispersive interactions, understanding the basic features of this new kind of of dissipative, directional dispersive interaction could provide powerful insights in how to usefully engineer complex open quantum systems.

In this work, we experimentally realize a dispersive type of quantum non-reciprocal interaction between a superconducting cavity and a transmon qubit.  This interaction is mediated by a complex set of cavity-magnon hybrid modes constructed from 3D niobium and copper cavities and a ferrimagnetic yttrium iron garnet crystal~\cite{Yingying_2021, Owens_2018,Tabuchi_2014,Xufeng_2014}. 
We study the mutual influence between the qubit and the cavity in terms of dispersive frequency shifts, dephasing and decay rates over parameter regimes with different degrees of non-reciprocity in situ tuned by an external magnetic field.  We highlight observations of non-reciprocal qubit-cavity frequency pulls and the applicability of generalized Onsager-type of relations upon time-reversal operations.  Importantly, despite the intractable complexity of the dissipative intermediary modes, we show that the qubit-cavity system dynamics can be captured effectively by a simple master-equation model, assuming a hierarchy of time scales.  Our work provides the first example of generalized dispersive interaction in open quantum systems. 

This article is organized as the following: In the next two sections, we introduce the experimental setup and present observations of dispersive qubit-cavity frequency shifts with varying degrees of non-reciprocity through Ramsey measurements.  Then we present a general master-equation model and extract the model parameters from measured free-evolution dynamics of the system starting from an initial cavity coherent state.  Finally, we test the validity of this model of dispersive non-reciprocity with additional experiments, including cavity susceptibility under continuous drive and cavity Fock state dynamics, followed by a brief conclusion and outlook.

\section{Device and System Setup}

\begin{figure}[tbp]
    \centering
    \includegraphics[scale=.29]{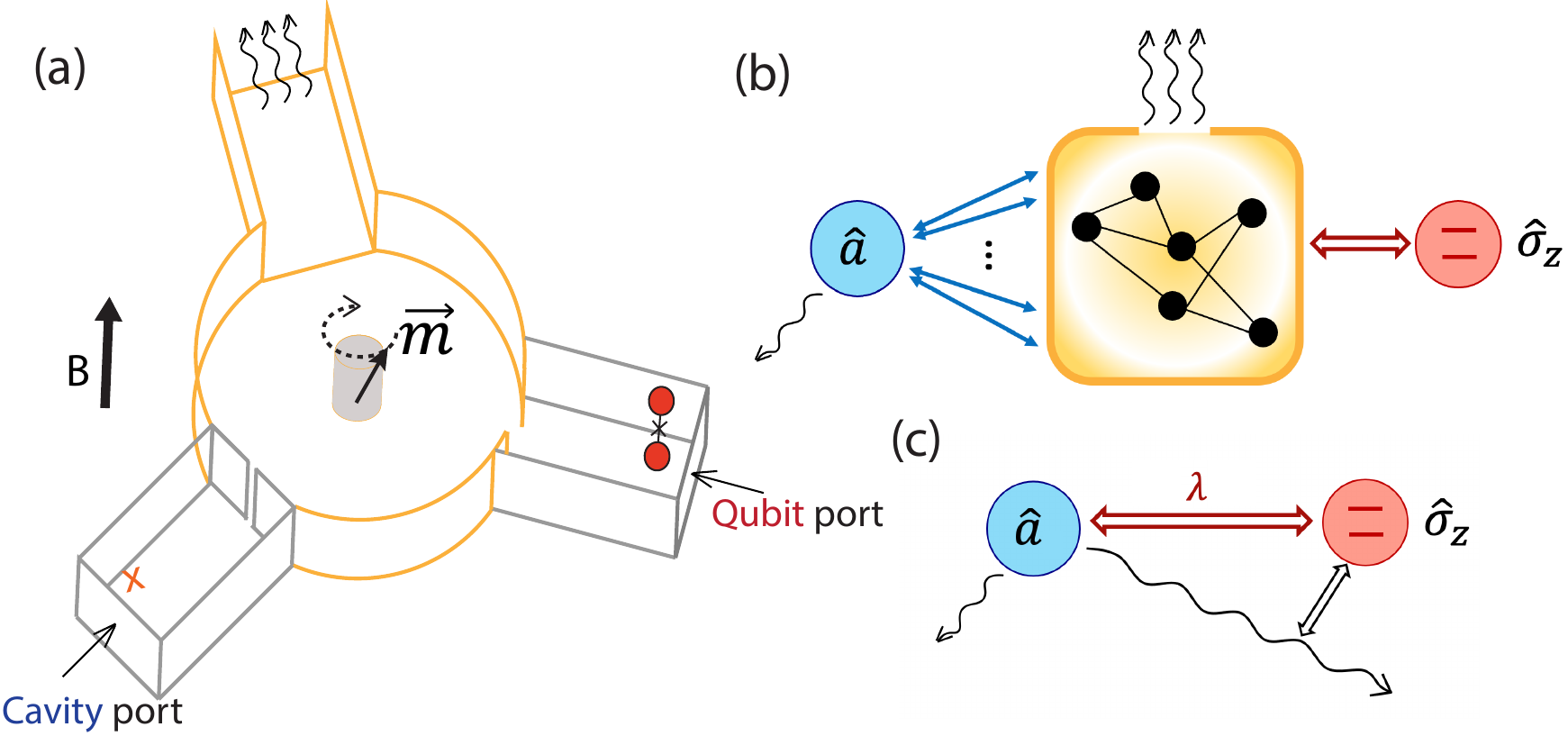}
    \caption{\textbf{Schematics of our integrated non-reciprocal device.} (a) A cartoon drawing of our device (not to scale), composed of a Cu waveguide Y-junction loaded with a YIG cylinder, two close-ended Nb rectangular waveguide segments with weakly-coupled drive ports, named as the ``Cavity port" and the ``Qubit port" respectively in the figure.  The lower left segment houses the TE201 cavity mode under study and an ancilla transmon (small orange cross) to facilitate readout of the cavity state.  The right segment houses the transmon qubit under study. The magnetic dipole in the YIG crystal are coupled to a series of microwave modes through its precession under external magnetic field. The upper waveguide segment of the Y-junction is impedance-matched to a transmission line for signal output. (b) A general schematic of the mode connectivity of system, where the cavity is exchange-coupled to many intermediary modes and the qubit is dispersively coupled to them.  The intermediary modes contains a large collective loss inherited from the open waveguide port. (c) Schematic representation of the qubit-cavity effective model, Eq.~(\ref{eq:master_eq}), with their reciprocal dispersive coupling $\lambda$ and a dispersive type of non-reciprocal dissipation operator $\Gamma  \mathcal{D} [  e ^{\frac{i \theta + \eta }{2} \hat \sigma_z  } 
    \hat a ]$.}
    \label{fig:device}
\end{figure}

Our experimental setup is shown in Fig.~\ref{fig:device}(a). 
A $\phi$-5.58 mm $\times$ 5.0 mm single-crystalline YIG cylinder is placed
at the center of a copper waveguide Y-junction, with external magnetic fields applied along its height (the [111] orientation of the YIG crystal).  The three sides of the Y-junction are connected, respectively, to (1) a niobium superconducting cavity (``the cavity"), (2) a superconducting transmon qubit (``the qubit") shielded in a closed-end niobium waveguide, and (3) a 50 $\Omega$ transmission line.  The collective spin precession inside the YIG crystal (magnon excitations) is hybridized with the electromagnetic modes in the vicinity of the waveguide Y-junction in a chirality-dependent manner, forming a series of photon-magnon polariton modes.   
The device geometry is reminiscent of a waveguide circulator. 
Indeed, at an external bias field of about $\pm 20$ mT, the ferrite-loaded Y-junction functions as a circulator that mediates directional microwave transmission between the cavity side and the transmon side with a bandwidth of a few hundred MHz~\cite{Yingying_2021}.  However, this device allows us to go beyond the special case of a canonical circulator, since the polariton modes can be tuned in-situ via the external magnetic field, yielding coupling channels with varying degrees of non-reciprocity.  More details of the YIG-cavity device platform and a modelling of the few-mode cavity polariton system without qubits can be found in an earlier article~\cite{Yingying_2021}.

Unlike in a practical circulator where broadband directional isolation is usually a first priority, the most important quality of our device is the low internal loss of the polariton modes. Therefore, we can describe the entire intermediary system (except for the cavity and the qubit themselves) as a network of coupled linear modes that share only one dominant decay channel, the 50 $\Omega$ transmission line, as shown in Fig.~\ref{fig:device}(b), regardless of the external field.  This allows us to treat the linear network, whose underlying details proves too challenging to characterize, as a single dissipative bath in mediating the effective interaction between the longer-lived qubit and cavity.  It is such a generalized quantum interaction, which encompasses both the standard dispersive interaction~\cite{Blais_2004} and the dissipative directional effects as represented by Fig.~\ref{fig:device}(c)~\cite{Yuxin_2023}, 
 that is the subject of our study. 
 
Integration of superconducting qubits in a hybrid quantum system with ferromagnetic magnons has been challenging because external magnetic fields degrade the qubit coherence.  Previous experiments used permanent magnets to provide strong local bias fields away from the qubit, achieving qubit lifetimes $T_1$ and $T_2$ of $<$1 $\mu$s~\cite{Tabuchi_2015,Dany_2017,Dany_2020} to a few $\mu$s~\cite{Owens_2018,Owens_2022}, but with limited to no tunability of external field.  In our experiment, we use the large electromagnet of the cryostat to apply a global magnetic field to the device, sufficient to fully reverse the directionality of the qubit-cavity coupling, while the qubit is shielded by the Meissener effect of the niobium waveguide and a layer of high-permeability foil.  We observe $T_1$ and $T_2$ on the order of a few $\mu$s (up to 10 $\mu$s) and unaffected by the applied field up to $\pm$0.1 T.

\begin{figure*}[tbp]
    \centering
    \includegraphics[scale=0.64]{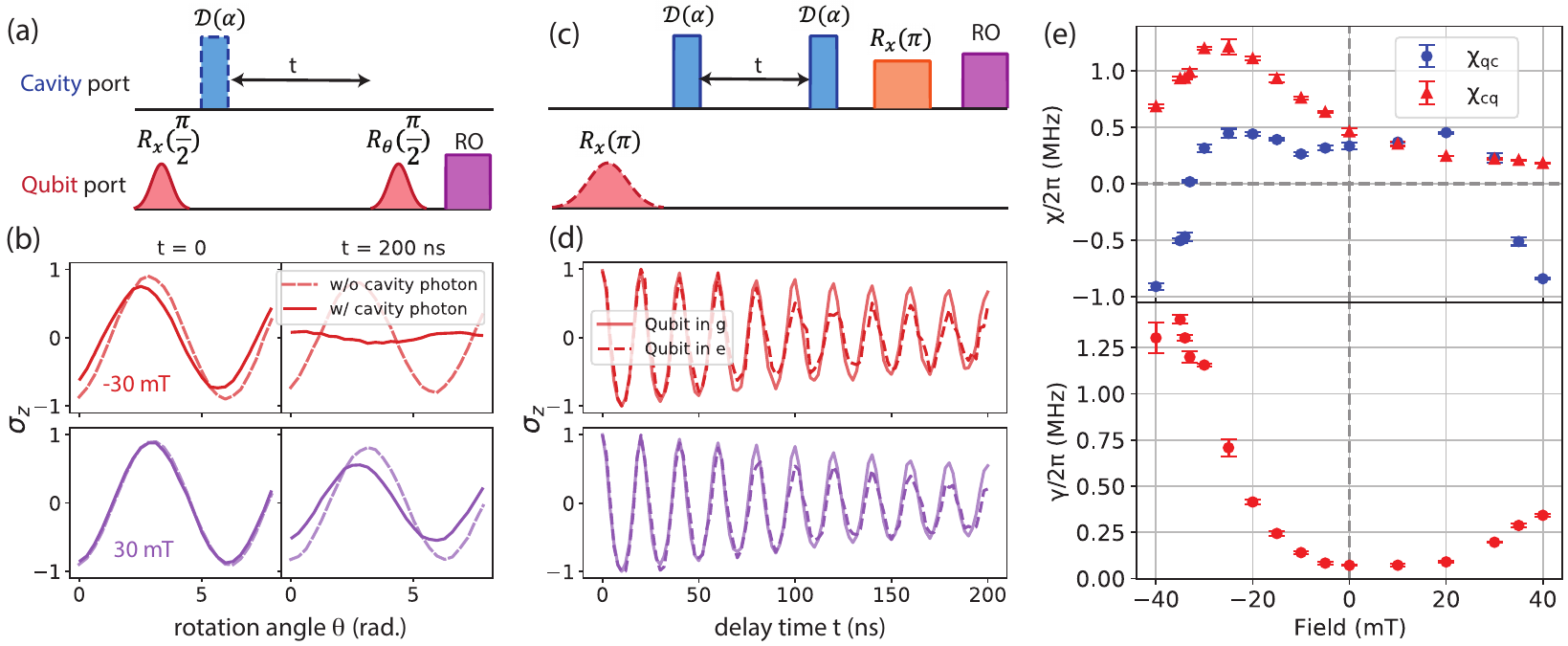}
    \caption{\textbf{Demonstration of non-reciprocal qubit-cavity frequency shifts.} (a) The pulse diagram and (b) example experimental results of a qubit Ramsey measurement under coherent cavity photon population at selective magnetic field (B = $\pm$30 mT). 
    The qubit coherence evolution is obtained by initializing the qubit in an equator state, and cavity in a coherent state with a displacement pulse $\mathcal{D}(\alpha)$. The system then evolves over a wait time $t$ followed by a second $\pi$/2 qubit rotation along a variable rotation angle $\theta$ and a readout of the qubit state.
    The qubit state against rotation angle $\theta$ yields a sinusoidal graph as plotted in (b), where the phase and amplitude can be extracted.
    The solid and dashed lines show the qubit Ramsey curves with and without cavity photons, respectively.
    (c) The pulse diagram and (d) the experimental result of the cavity photon Ramsey measurement with a relatively long ancilla excitation pulse (orange $R_x(\pi)$) to obtain the qubit dependant cavity frequency, where the solid (dashed) lines correspond to the qubit in $\ket{g}$ ($\ket{e}$).  
    (e) The upper panel plots both measured $\chi$s under different external magnetic fields showing variable non-reciprocity between the cavity and the qubit. $\chi_\mathrm{qc}$ is near-symmetric across the external field, as apparent in (d), where the -30 mT and 30 mT curves are similar to each other. $\chi_\mathrm{cq}$ is clearly asymmetric, where the positive and negative field results are distinctly different, as shown in (b). The lower panel is the qubit dephasing rate under different external magnetic fields.
    \label{fig:ramseys}
    }
\end{figure*}

The cavity in our experiment, whose mode of interest is the TE201 mode at 10.809 GHz, is evanescently coupled to the Y-junction through an aperture, giving a cavity lifetime of 40-100 ns (dependent on the field and the qubit state) dominated by coupling to the polariton modes in the Y-junction.  The cavity can be driven from a weakly-coupled cavity port as labeled in Fig.~\ref{fig:device}(a).  An additional transmon (marked as orange $\times$) is installed inside the cavity for readout of the cavity state, which will not participate in the quantum dynamics under study and will be referred to as the ancilla, to be distinguished from the qubit.  On the other hand, the qubit can be driven from a weakly-coupled qubit port as labeled in Fig.~\ref{fig:device}(a).  The qubit at 9.141 GHz is Purcell protected by effectively a buffer cavity mode formed by a modest constriction slot between the waveguide segment and the Y-junction. This extra buffer mode is sufficiently short-lived (lifetime in the range of 5-25 ns) to be treated as part of the dissipative bath rather than a quantum object, but it provides an impedance environment to boost the interaction between the qubit and the cavity.  Throughout the main text, we will present data from this device.  We have also carried out the experiments on a second device with similar cavity parameters but different transmon parameters, and the results are included in the supplementary materials.

\section{Observation and control of non-reciprocal frequency shifts}

As the qubit and cavity are detuned, we expect the mediated interaction to be dispersive. 
To characterize the phenomenon of dispersive non-reciprocity between the qubit and the cavity, we compare the qubit frequency shift per cavity photon, labeled $\chi_\mathrm{cq}$, with cavity frequency shift in response to the qubit excitation, labeled $\chi_\mathrm{qc}$.  A closed quantum system is reciprocal, where the dispersive frequency pull per excitation in both directions are guaranteed to be equal, i.e.~$\chi_\mathrm{cq}=\chi_\mathrm{qc}$. Violation of this relation is a sign of non-reciprocity.

To measure $\chi_\mathrm{cq}$, we perform Ramsey experiments on the qubit with and without cavity photons.  By comparing these measurements, we can extract the accumulated extra qubit phase shift $\phi$ caused by cavity photons over a finite time window of $t = 200$ ns, as shown in Fig.~\ref{fig:ramseys}(a).  The cavity is initialized in a coherent state with an initial mean photon number of $\bar{n}_0\approx 3$.   The choice of $\bar{n}_0$ and $t$ is motivated by optimizing signal-to-noise ratio while keeping the cavity well within the linear regime.    
The cavity photon number undergoes free decay over time, and hence the instantaneous qubit frequency also varies over time.  Assuming the dispersive frequency shift is proportional to photon numbers (as is the case in closed-system circuit QED), we define the cavity-to-qubit dispersive shift $\chi_\mathrm{cq} = \phi/(t\cdot\bar{n}_\mathrm{avg})$, where $\bar{n}_\mathrm{avg}$ is the time-averaged cavity photon number during the integration time $t$.
Example results of the qubit state versus rotation angle $\theta$ of the second Ramsey $\pi/2$ pulse are shown in Fig.~\ref{fig:ramseys}(b) (solid lines), which can be fit to a simple sinusoid to accurately extract the amplitude and phase of the qubit coherence function $\langle \sigma_- (t)\rangle$ at fixed time $t$.  By comparing $\langle \sigma_- (t)\rangle$ to the reference qubit state $\langle \sigma^0_- (t)\rangle$ measured without cavity photons (dashed lines) at the same $t$, we can extract the photon-induced phase shift $\phi$ and decoherence factor $\zeta$ at time $t$ from:
$\langle \sigma_-\rangle / \langle \sigma^0_-\rangle = \zeta e^{i\phi}$.
The time-averaged cavity photon number $\bar{n}_\mathrm{avg}$ is calibrated using a separate Ramsey experiment of the ancilla inside the cavity for the same time window, which allows for normalization of $\chi_\mathrm{cq}$ as a per-photon quantity.  (See Supplementary Materials~\ref{subsec:photon_number_cal} for more detailed discussions of the calibration.)  
The extracted decoherence factor $\zeta$ informs the cumulative loss of qubit coherence due to photon shot noise over the time window $t$, which can be similarly converted to a qubit dephasing rate per photon, $\gamma = -\ln(\zeta)/(t\cdot \bar{n}_\mathrm{avg})$.

On the other hand, $\chi_\mathrm{qc}$ is measured with a cavity Ramsey protocol with the qubit in its ground ($\ket{g}$ or $\ket{\uparrow}$) or excited ($\ket{e}$ or $\ket{\downarrow}$) state. 
The cavity Ramsey sequence is composed of two cavity displacement pulses with a wait time $t$ in between, as shown in Fig.~\ref{fig:ramseys}(c).  
We use the dispersively-coupled ancilla inside the cavity to read out the cavity state.  This readout, inspired by photon number measurements in the strong dispersive regime~\cite{schuster2007_number_splitting, kirchmair2013_Kerr_revival}, is implemented by a relatively long (spectrally narrow) ancilla excitation pulse,  followed by reading out the state of the ancilla.  Although our system is not in the number-resolved regime, the ancilla pulse excites the ancilla with decreasing efficiency at increasing photon numbers.  Therefore, the measured ancilla $\langle \sigma_z \rangle$ provides a monotonic proxy for the cavity photon number (even though the relationship is not linear), which is sufficient for an accurate measurement of the free-evolution frequency of the cavity.  Example results of the ancilla state over $t$, representing the oscillation and decay of the cavity coherent state 
is plotted in Fig.~\ref{fig:ramseys}(d), with the solid and dashed lines corresponding to the qubit in $\ket{g}$ and $\ket{e}$ state respectively.  
Since the qubit lifetime $T_1 \gg t$, the qubit state does not change over the measurement window to a good approximation, as confirmed by the constant cavity oscillation frequency in this measurement.  The extracted cavity frequency $\omega_{g}$ and $\omega_{e}$ gives $\chi_\mathrm{qc} = \omega_{g} - \omega_{e}$.  

The two types of Ramsey measurements have been carried out for various external magnetic fields, and the extracted dispersive shifts in both directions, $\chi_\mathrm{qc}$ and $\chi_\mathrm{cq}$, are plotted in the top panel of Fig.~\ref{fig:ramseys}(e).  The magnetic field serves as a control knob to vary the complex bath-mediated qubit-cavity interaction.  
While it is not surprising that non-reciprocity exists
in the presence of magnetic field, our experiment presents an unambiguous signature of non-reciprocity in the dispersive regime, $\chi_\mathrm{qc}\neq \chi_\mathrm{cq}$,  
and demonstrates in situ control over the degree of such non-reciprocity, e.g.~ranging from approximately reciprocal near zero field to strongly non-reciprocal at high negative fields.  Moreover, we note a few non-trivial features in the magnetic field dependence.  First, $\chi_\mathrm{qc}$ shows symmetry with respect to $B$ while $\chi_\mathrm{cq}$ obeys no such symmetry. 
The former is a result of an Onsager-type of eigenvalue relationship for linear non-Hermitian Hamiltonian, which arises from microscopic time-reversal symmetry even in the presence of external field.  On the other hand, $\chi_\mathrm{cq}$ and the photon-induced dephasing rate $\gamma$ (the bottom panel of Fig.~\ref{fig:ramseys}(e)) are governed by different dynamics. 
Second, the non-reciprocity at zero field is small but definitely non-zero within experimental uncertainties.  While many experiments have realized magnetless non-reciprocity by engineering synthetic flux~\cite{Nicholas_2014,Chapman_2017}, our data at zero field demonstrates an important theoretical aspect of quantum non-reciprocity: The dissipative interaction between quantum subsystems can be non-reciprocal without real or synthetic magnetic field~\cite{Yuxin_2023}.

\section{Effective non-reciprocal model in the dispersive regime}

While the phenomenological observables $\chi_\mathrm{qc}$ and $\chi_\mathrm{cq}$ highlight a distinctive aspect of the effective qubit-cavity dispersive interaction, they do not a priori fully characterize the general dynamics of the qubit-cavity system.  Given an arbitrary initial state of this qubit-cavity system, how can we model the system to fully predict its time evolution?  Of course, if we know the full details of all relevant intermediary modes (Fig.~\ref{fig:device}(b)), including their mode frequencies, decay rates, internal and external coupling rates, one could in principle solve the dynamics of the expanded system.  However, not only is this approach computationally expensive and intuitively opaque, it is often unrealistic to extract detailed knowledge of a highly-dissipative multi-mode system. 
On the other hand, if we assume the system is in a regime where it is valid to adiabatically eliminate the intermediary modes and the dispersive approximation holds, we can derive a simple effective Markovian master equation only involving the cavity $\hat{a}$ and the qubit:
\begin{equation}
\begin{aligned}
\partial_t \hat \rho 
=   -i  \left[ 
\Delta_c \hat a ^\dag \hat a  
+ \frac{ \lambda  }{2} \hat \sigma_z  \hat a ^\dag \hat a , \hat \rho   \right]\\
	+ \kappa 
	\mathcal{D} \left[   \hat a  \right]  \hat \rho  
	+ \Gamma  
	\mathcal{D} \left[  
 e ^{\frac{i \theta + \eta }{2}
 \hat \sigma_z  }
   \hat a  \right]  \hat \rho  
\end{aligned} 
\label{eq:master_eq}
\end{equation}
which is written in the rotating frame of both the qubit and a reference frequency of the cavity (see Supplementary Materials~\ref{subsec:adia.elim.multi}  for details), and can be fully specified via 6 independent real parameters.  Here, $\lambda $ is the conventional reciprocal dispersive shift between the cavity and the qubit, $\kappa$ is the conventional decay rate of the cavity mode unrelated to the qubit, 
and $\Gamma \mathcal{D} \left[  e ^{\frac{i \theta + \eta }{2} }   \hat \sigma_z \hat a  \right]  \hat \rho$ describes a form of non-linear collective dissipation with $\Gamma $ the dissipation rate.  Its nonlocal jump operator $ e ^{\frac{i \theta + \eta }{2} \hat \sigma_z  } \hat a  $ provides a uni-directional frequency pull on the qubit through imaginary part of the exponent $\theta $~\cite{Yuxin_2023}, while its real part $\eta$ couples cavity decay to qubit dephasing.  
We note that $\hat \sigma_z$ is a conserved quantity in Eq.~\eqref{eq:master_eq}, as the coupling between the qubit and the bath is in the far-detuned limit.  We have not included here the intrinsic dissipation of the qubit that would give rise to additional qubit-only dephasing and relaxation dissipators.  We stress that this effective model allows description of arbitrary quantum dynamics in the qubit-cavity Hilbert space, and is not limited to semi-classical equations of motion.  
Nonetheless, we can uniquely specify the model from a convenient set of measurements where the cavity dynamics can be described by semi-classical mode amplitudes and fluctuations, and later apply this model to other scenarios for verification.

\begin{figure}[tbp]
    \centering
    \includegraphics[scale=.78]{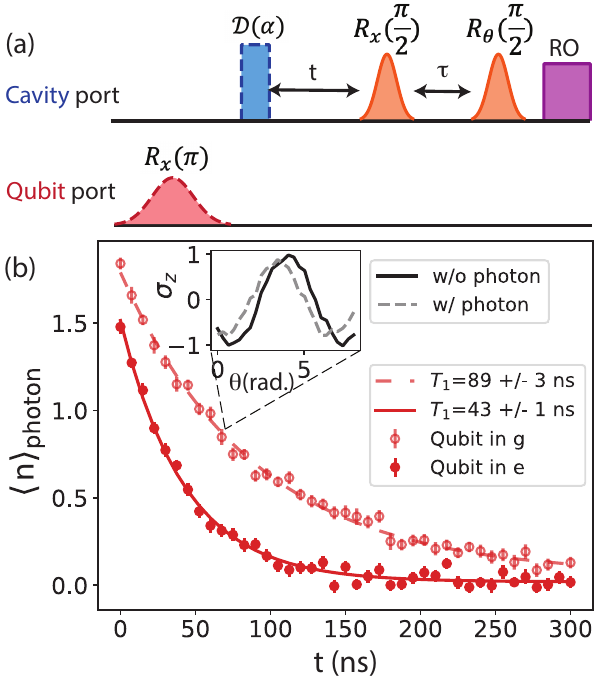}
    \caption{\textbf{Qubit state dependent cavity $T_1$ measurements.} (a) The pulse diagram and (b) the experimental result of cavity decay rate measurement, 
    with qubit prepared in $\ket{g}$ or $\ket{e}$. The cavity is prepared in a coherent state with a displacement pulse $D(\alpha)$, and the phase shift on the ancilla is measured through a Ramsey measurement over the fixed time window $\tau$. This phase shift can be related to the cavity photon number through the dispersive shift.
    This photon number over the variable time $t$ gives the cavity decay rate (cavity $T_1$) while the qubit is in $\ket{g}$ or $\ket{e}$. 
    }    
    \label{fig:cavT1s}
\end{figure}

To motivate our measurement protocol to determine the model parameters, it is worth first discussing the physical meaning of the dynamics generated by Eq.~\eqref{eq:master_eq}: the cavity now experiences a qubit-state-dependent frequency detuning and decay rate, whereas the qubit undergoes a time-dependent, cavity-photon-induced phase shift and dephasing effect. From the master equation in Eq.~\eqref{eq:master_eq}, it is straightforward to see that the average cavity amplitude 
$\bar{a} _{\sigma_z } (t)$ is now contingent on qubit state 
$| \sigma_z \rangle$ 
($\sigma_z = $ $\uparrow, \downarrow $), whose effective complex frequency 
$\mathcal {E} _{ \sigma_z } $, as defined via the cavity equation of motion 
$i ( d \bar{a}  _{\sigma_z }/{dt} ) 
=  \mathcal {E} _{ \sigma_z }   
\bar{a}  _{\sigma_z }
$, is given by   
\begin{align}
\label{eq:def_cav1_freq}
\mathcal {E} _{ \sigma_z }  = 
\Delta_c 
+ \frac{\lambda }{2} \sigma_z
-i \frac{ \kappa 
+ \Gamma  
e ^{  \eta \sigma_z  } 
  }{2}, 
\end{align}

Therefore, measurements of the cavity frequency $\omega_g$, $\omega_e$ and decay rates $\kappa_g$, $\kappa_e$ for the separate scenarios when the qubit is initialized either in its $\ket{g}$ or $\ket{e}$ state would uniquely determine $\Delta_c$, $\lambda$, and provide two constraining equations for the four parameters ($\kappa$, $\Gamma$, $\eta$, and $\theta$) in the dissipation terms in Eq.~(\ref{eq:master_eq}).

Two more constraints can be established using the qubit Ramsey experiment with the cavity initialized in a coherent state (at $t=0$). 
From Eq.~\eqref{eq:master_eq}, the qubit coherence obeys the equation of motion (see Supplementary Materials~\ref{subsec:simulation_detail} for derivations): 
\begin{align}
\label{eq:eom_qb_coh}
&   \frac{ d \langle 
\hat \sigma_-  (t) \rangle }
{  dt } 
= 
    \left[ -  i \lambda  
    +\Gamma  (e ^{ i \theta}  
    - \cosh \eta )
    \right] 
    \bar{a} _{\uparrow} 
    \bar{a} ^*_{\downarrow} 
    \left \langle \hat \sigma _ - (t) 
\right \rangle 
    . 
\end{align}
where the cavity amplitudes
$\bar{a}_{\uparrow/\downarrow}(t)$
can be solved from Eq.~\eqref{eq:def_cav1_freq}. 
In the long-time limit that $t_f \gg 1/(\kappa+\Gamma)$, the integrated change in qubit coherence function is connected to the system parameters via 
\begin{align}
\ln \frac{ \left\langle\hat{\sigma}_{-}\left(t_{f}\right)\right\rangle}{\left\langle\hat{\sigma}_{-}(0)\right\rangle} 
\simeq 
\bar{n}_0  
\frac{
-i\lambda +  \Gamma (e ^{ i \theta}  
- \cosh \eta ) }
{i\lambda +\kappa +\Gamma \cosh \eta 
}
.  
\label{eq:qb.ramsey.tf}
\end{align}
where $\bar{n}_0$ is the initial cavity photon number that can be calibrated independently.

Measurements of the six real observables (i.e.~three complex numbers) on the LHS of equations \eqref{eq:def_cav1_freq} and \eqref{eq:qb.ramsey.tf} together provide sufficient constraints to uniquely determine all the parameters in the master equation Eq.~\eqref{eq:master_eq}. 
Note that so far in this derivation of Eq.~\eqref{eq:master_eq} we have neglected intrinsic decoherence of the qubit for simplicity, which in experiments is calibrated out by performing differential measurements with or without the cavity drives.  Most of these experiments have already been described in Section III, including the qubit Ramsey experiment in measuring the qubit coherence function in the presence of cavity photons (where we use data at $t_f=700$~ns instead of 200~ns in this analysis), and the cavity Ramsey experiment in measuring the cavity frequency, $\omega_{g, e}$. 
Here we discuss an additional experiment to accurately extract the cavity decay rates 
$\kappa_{g,e}$, or equivalently the cavity $T_1$ times.

The main challenge of the cavity $T_1$ experiment is to implement a reliable readout scheme that detects the mean cavity photon number in the range of 0.1-10 in a linear fashion.  Direct heterodyne detection of cavity photons requires much larger photon numbers for reasonable measurement time, which will be affected by higher-order spurious nonlinearities.  Spectroscopic probe of the dispersively-shifted ancilla frequency, as done in Section III, is complicated by the fast decay of the cavity photons during the relatively slow spectroscopy pulses.  
Here, we map the cavity photon number to the ancilla's phase shift $\phi$ over a sliding time window of fixed length ($\tau=100$ ns, as shown in Fig.~\ref{fig:cavT1s}(a)), which is measured in a Ramsey sequence of the ancilla.  Although the instantaneous cavity photon number changes substantially during the time window $\tau$, the cumulative phase shift $\phi_a$ can be used to infer the average photon number during $\tau$, $\bar{n}_\mathrm{avg}=\frac{\phi_a}{\tau\chi_a}$, with $\chi_a$ the dispersive shift between the cavity and the ancilla.  We find that $\bar{n}_\mathrm{avg}$ can be fit well to an exponential decay as the Ramsey window slides in time $t$ for both the qubit in $\ket{g}$ and in $\ket{e}$ (Fig.~\ref{fig:cavT1s}(b)), which gives the cavity decay rates for the respective cases. In our experiment, we observe that our cavity decay rates fluctuate over the time scale of hours to days, possibly caused by changing configurations of trapped vortices, but the difference, $\kappa_g-\kappa_e$, tends to be mostly stable. 

\begin{figure}[tbp]
    \centering
    \includegraphics[scale=.7]{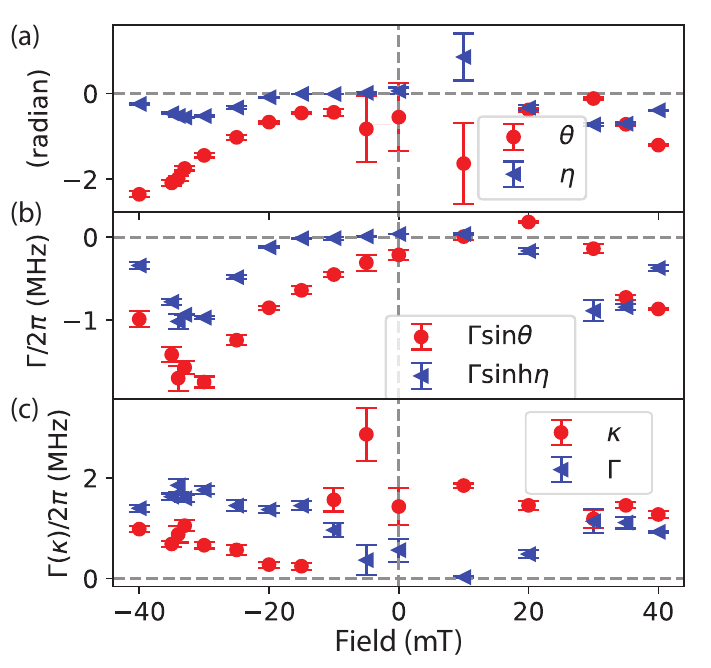}
    \caption{\textbf{Extracted master equation parameter.} With all of the presented measurements (qubit Ramsey, cavity Ramsey and cavity $T_1$), 
    the dissipation related parameters for the master equation Eq.~(\ref{eq:master_eq}) are plotted, see Eq.~\eqref{eq:def_cav1_freq}-\eqref{eq:qb.ramsey.tf} for deriving details. 
    }    
    \label{fig:params}
\end{figure}

Combining all of the experiments discussed so far, we can uniquely determine the parameters of Eq.~\eqref{eq:master_eq} that describes the full quantum dynamics of the qubit-cavity system.  The applied magnetic field $B$ provides an in situ tuning knob that allows us to access many distinct instances of the model, and each can be determined independently.  In Fig.~\ref{fig:cavT1s}(c), we show four of the six extracted model parameters:
$\theta$, $\eta$, $\Gamma$ and $\kappa$ as a function of field.  The dispersive interaction $\lambda$ is equivalent to $\chi_{qc}$ as plotted in Fig.~\ref{fig:ramseys}(e), and the relatively trivial cavity detuning $\Delta_c$ can be found in the Supplementary Materials~\ref{subsec:params}.  In Fig.~\ref{fig:cavT1s}(c), we further plot two particular parameter combinations in the non-reciprocal dissipator $\Gamma\mathcal{D}[e^{\frac{i\theta+\eta}{2}\hat\sigma_z}\hat{a}]$ that are directly linked to experimental observables: $\Gamma\sin\theta$ reflects
the non-reciprocal frequency shift and $\Gamma\sinh\eta$ corresponds to the qubit-state-dependent cavity decay rate. 

How the magnetic field controls each of the model parameters depends on complex details of the intermediary bath modes, which is not a focus of our study.  Nevertheless, we observe
in Fig.~\ref{fig:cavT1s}(c) that
$\Gamma\sinh\eta$ at $\pm B$ are equal, and the symmetry also holds for $\chi_\mathrm{qc}$ in Fig.~\ref{fig:ramseys}(e).  This is no coincidence, but the consequence of microscopic symmetry requirements.  In a linear systems with micro-reversibility, the Onsager-Casimir relation~\cite{casimir_1945} requires that the full scattering matrix $S$ satisfy $S(-B) = S^\mathrm{T}(B)$. Since our original multimode system dynamics conserves the qubit Pauli operator $\hat \sigma_z$, the effective model parameters also satisfy corresponding relations, i.e.~$\left\langle\hat{a}_{\sigma_z}\right\rangle(B) = \left\langle\hat{a}_{\sigma_z}\right\rangle(-B)$, which by referring to Eq.~\eqref{eq:def_cav1_freq}, directly lead to the symmetry of $\lambda $ (i.e.~$\chi_\mathrm{qc}$), $\Delta_c $, $\Gamma \sinh  \eta$, 
$\kappa  + \Gamma  \cosh \eta $ (see Supplementary Materials~\ref{subsec:Onsager.disp} for details).  This constraint lets us directly understand the symmetry shown on figures mentioned earlier.  Note that 
an analogous theoretical argument predicts that $\kappa$ should be symmetric in $B$, something that is apparently violated by the data in Fig.~\ref{fig:cavT1s}(c).  This might be explained by the presence of other long-lived modes in the system (such as a different cavity mode or a magnon mode) that may get accidentally excited and also have couplings to the qubit, which makes the calculation of $\Gamma$ and thus $\kappa$ less accurate.

\section{Verification of the qubit-cavity dynamics}
\label{main_sec:verif.drives}

\begin{figure}[tbp]
    \centering
    \includegraphics[scale=.75]{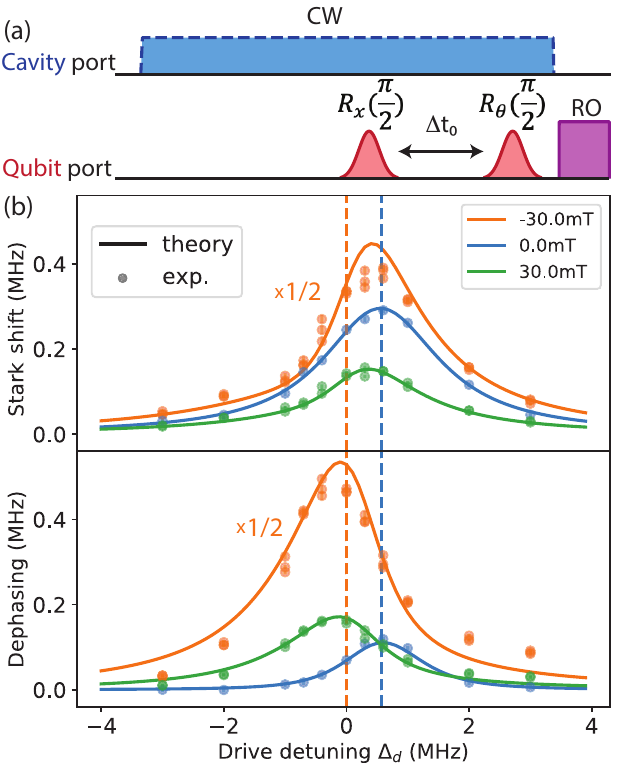}
    \caption{\textbf{Parameter-free verification of the master equation model with continuous cavity drive.} (a) The pulse diagram and (b) example of the theory prediction (solid line) and experimental result (dot) of qubit Ramsey measurement under steady state CW (5 $\mu$s) cavity drive at selective fields (constant drive amplitude across detunings). The Ramsay is performed with a fixed time window at the end of the CW tone. At each CW detuning the Ramsay data gives the Stark shift and dephasing rate. The reslut at $-30$ mT is scaled by 1/2 for better visualization. 
    The blue dashed line indicates the cavity frequency $\Delta_c$ at 0 mT, and the 0 mT Stark shift and dephasing peaks correspond well with this frequency, as expected. The orange dashed dot line indicates $\Delta_c$ at $\pm$30 mT.  Here the peak frequency in Stark shift and dephasing rates are distinctly different from each other. See Supplementary Materials~\ref{subsec:cw} for more data at different fields. 
    }
    \label{fig:cw}
\end{figure}

At this stage, we have used a set of experimental measurements to determine the parameters of the general master equation model in Eq.~\eqref{eq:master_eq} that should describe a generic non-reciprocal, dissipative dispersive qubit-cavity interaction.  Of course, this extraction of parameters does not by itself show the validity or utility of our model.  
We address these issues here:  we now use our fully constrained model to make predictions for {\it new} experiments (with different initial states and/or drives), and compare these directly against experiment.  

\begin{figure*}[tbp]
    \centering
    \includegraphics[scale=.7]{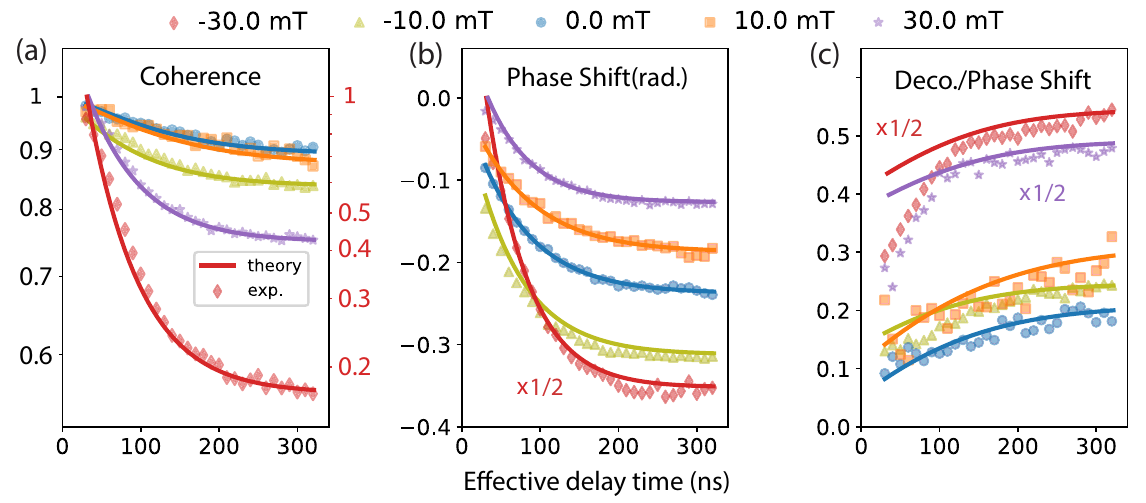}
     \caption{\textbf{Experiment-theory comparison of the transient dynamics.} Time domain values of (a) qubit coherence ($\zeta$), (b) accumulated phase shift ($\phi$) and (c) the ratio beetween qubit decoherence and phase shift ($\ln(\zeta)/\phi$) of qubit Ramsey measurement results at $-30$ mT (red), $-10$ mT (yellow), $0$ mT (blue), 10 mT (orange) and 30 mT (purple). The cavity is initialized in a coherent state for all data. The scatter dots are experimental results and solid lines are theory predictions from the master equation Eq.~\eqref{eq:master_eq}.
     The $-30 mT$ data in the coherence panel is plotted according to the red y scale on the right side of the panel, the $-30$ mT data in phase shift panel and $\pm$30 mT data in the decoherence/phase shift panel are scaled by 1/2 for better relative visual scaling. Experimental dots start at 30 ns considering the 60 ns qubit pulse right before readout pulse. The theory curve for $\pm$30 mT has been shifted to the right by an amount 33 ns and 31 ns, where the shift amount is a fit parameter, to capture the evolving time of intermediary modes, which is considered to be instantaneous in the model, providing a good match to the experimental data.
     }    
    \label{fig:time}
\end{figure*} 

In the first verification experiment, we investigate the qubit response to continuous cavity drive.  Here, we apply a 
continuous wave (CW) drive to the cavity at constant amplitude and varying frequencies, and measure the resultant ac Stark shift and photon shot-noise-induced dephasing rate on the qubit using a Ramsey sequence (Fig.~\ref{fig:cw}). 
Note that the cavity is stabilized to a steady state during the Ramsey protocol as opposed to undergoing free decay in the experiments presented in Sec.~III.
The model prediction of the driven system, which can be solved after appending a drive term $\epsilon (\hat{a}^\dagger e^{- i\Delta _{ d } t} + h.c.)$ to the Hamiltonian in Eq.~\eqref{eq:master_eq}, agrees quite well with the experimental data with {\it no free parameters} (see Supplementary Materials~\ref{subsec:validation_cavdr} for details).
We observe that the Stark shift and dephasing rate display distinctly different peak frequency at selected fields (e.g.~$30$ mT and $-30$ mT as shown in Fig.~\ref{fig:cw}(b)), which is unexpected for a traditional dispersively-coupled qubit cavity system and therefore is a signature of the non-reciprocal interaction in our system.  This non-trivial feature in the frequency and lineshape is well captured by our model.
In comparison, at 0 mT, when the system is close to reciprocal, the Stark shift and decay rate peak at similar frequency.

In another test of the master equation model, we investigate the time-domain evolution of the qubit in the presence of cavity photons.  The experimental protocol is the same as in Fig.~\ref{fig:ramseys}(a), where the cavity is initialized in a coherent state and the qubit is initialized in an equator state.  In Fig.~\ref{fig:time}(a, b) we show the qubit coherence factor $\zeta$ and the cumulative phase shift $\phi$ as a function of time $t$ for a range of external fields. 
The slopes of the coherence factor (on log scale) and the phase shift on these plots correspond to the instantaneous photon-shot-noise dephasing rate and a.~c.~Stark shift, and the decrease of slopes over time indicates the continuous decay of cavity photon numbers.  A non-trivial feature to be noted is that the dephasing and frequency shift effects decay on slightly different time scales, which can be seen more clearly from their ratio, 
$\ln(\zeta)/\phi$, which is not a constant over time (Fig.~\ref{fig:time}(c)). 
The time evolution based on the master equation model can be solved as shown in the Supplementary Materials (Eq.~\eqref{eq:coherence_evolution}) and are plotted as solid curves for comparison.  Since the model is applicable on time scales where all fast bath degrees of freedom can be adiabatically eliminated, discrepancies are expected on short time scales.  This is especially true at higher magnetic fields when the buffer cavity enclosing the qubit has longer lifetimes ($\gtrsim20$ ns).  The model captures the time-domain data very well at low fields (e.g.~0 mT and $\pm10$ mT).  At higher fields, we find that the model can describe the temporal evolution reasonably if one allows for an empirical time offset as a fit parameter, which is 33 ns and 31 ns for $\pm30$ mT in Fig.~\ref{fig:time}.  This ad hoc modification to the model can be understood as some ``turn-on" time allowing for the bath (which is not infinitely fast) to reach the quasi-steady state set by the initial condition of the qubit-cavity system. 
Fig.~\ref{fig:time}(c) provides a more sensitive test of the model, which correctly captures the time-varying ratio of dephasing and phase shift effect, although the discrepancies at short times and higher fields become apparent due to the lack of separation of time scales.

\begin{figure}[tbp]
    \centering
    \includegraphics[scale=.65]{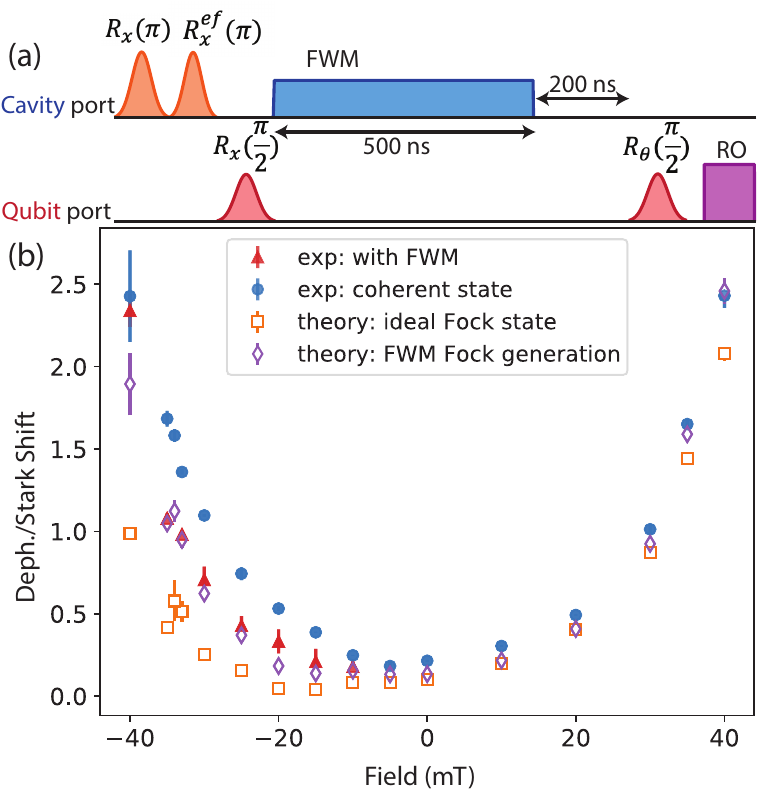}
    \caption{\textbf{Parameter-free verification of the master equation with single photon generation.} 
    (a) The pulse diagram of qubit Ramsey experiment in the presence of a single photon generated by a four-wave-mixing (FWM) pulse. 
    After initializing the ancilla in $\ket{f}$, a $\ket{f0}-\ket{g1}$ FWM pulse is applied to convert the ancilla $\ket{f}$ state to a cavity photon. 
    (b) The experimental result of the qubit dephasing/Stark shift ratio from the $\ket{f0}-\ket{g1}$ FWM measurement (red triangle) and coherent cavity state measurement (blue circle), compared to theory predictions with different initial states: ideal single photon Fock state (orange square), FWM Fock state generation mimicking the experimental setups (purple diamond), which considers the initial thermal population of the ancilla and pulse infidelity thus set ancilla state initializing at 85\% in $\ket{f}$ state, and the ancilla $T_1$ decay during the FWM pulse. 
    }
    \label{fig:fwm}
\end{figure}

Lastly, we apply the non-reciprocal master equation Eq.~\eqref{eq:master_eq} to scenarios with non-Gaussian cavity states.  Specifically, we consider cavity-qubit dynamics where the cavity is initialized in a single-photon Fock state.   
Here we can no longer understand the system via semi-classical equations of motion for mode amplitudes (e.g.~as in Eq.~\eqref{eq:def_cav1_freq}-\eqref{eq:qb.ramsey.tf}), but Eq.~\eqref{eq:master_eq} remains valid and provides direct insights to the non-reciprocal qubit-cavity interaction.  An interesting special case is $\eta=0$, where the dissipator $\Gamma\mathcal{D}[e^{\frac{i\theta}{2}\hat{\sigma}_z}\hat{a}]$ implements a unitary gate $e^{\frac{i\theta}{2}\hat{\sigma}_z}$ on the qubit when the photon escapes the cavity.   Therefore, if the cavity does not lose the photon via other channels (i.e.~$\kappa=0$), the qubit will receive a deterministic phase shift of $\theta$ after $t\gg 1/\Gamma$ without incurring photon-induced dephasing.  A practical application of this deterministic phase shift is microwave single-photon detection as was implemented in Ref.~\cite{Besse_2018,Kono_2018}.  These experiments used a conventional circulator to enforce directionality, and were understood as the interaction between a qubit and a travelling photon in a transmission line.  Our model effectively describes the interaction between the detector qubit and the source cavity of the photons, and is further generalized to allow the non-reciprocity of the interaction channel to be varied continuously.  

Our experimental platform, aided by the ancilla inside the cavity, allows us to generate single photons and investigate their interaction with the qubit over a range of model parameters by varying magnetic fields.   
A cavity Fock state $\ket{1}$ can be generated by first preparing the ancilla transmon in the $\ket{f}$ state (its second excited state), and applying a strong drive to induce the four-wave-mixing (FWM) $\ket{f0}-\ket{g1}$ sideband transition, which converts the double excitations in the ancilla to a single cavity photon.  Experimentally, our FWM transition rate is limited to 0.6 MHz, which, compared to cavity total decay rate of 1.7 MHz, is not fast enough to instantaneously initialize the cavity in $\ket{1}$ before the dissipative process takes place.  Nevertheless, in the limit where the ancilla's $\ket{f}$ state is perfectly prepared and long-lived, under a sufficiently long FWM drive, there will be one and only one photon generated and emitted from the cavity, effectively realizing single-photon dynamics in our qubit-cavity system. 
We implement this approach of single-photon generation with a pulse sequence as shown in Fig.~\ref{fig:fwm}(a), and measure the long-time ($t=700$ ns) cumulative phase shift $\phi$ and dephasing $\ln(\zeta)$ of the qubit similarly to in previous coherent-state-based  experiments.  When the bath modes of the device function approximately like a circulator in the direction from the cavity to the qubit (near $B\approx -20$ mT), one can expect suppressed qubit dephasing but non-zero phase shift for a single photon.  Therefore, the suppressed ratio of $\ln{(\zeta)}/\phi$ is a distinctive signature of the Fock state-qubit interaction compared to a coherent state-qubit interaction.  This ratio can be solved from Eq.~\eqref{eq:master_eq} (See Supplementary Materials~\ref{subsec:Fock_FWM_Fock}) and is appreciably smaller than the coherent-state case whenever $\Gamma\sin\theta \lesssim \Gamma \sinh\eta, \kappa$, which corresponds to most of the negative field range in our system (c.f.~Fig.~\ref{fig:params}(b)).  
In Fig.~\ref{fig:fwm}(b), indeed we show that $\ln{(\zeta)}/\phi$ is smaller for the Fock-state induced dynamics for this field range.  
At positive fields, both phase shift and dephasing are too small to be measured accurately in the single photon limit.  The experimentally measured dephasing factor in the Fock-state experiment, as reflected in the $\ln{(\zeta)}/\phi$ ratio (red data points) is larger than Eq.~\eqref{eq:master_eq} prediction assuming a perfect initial Fock state (orange).  The discrepancy is caused by two factors: one is limited conversion efficiency (about 89\%) of the $\ket{f0}-\ket{g1}$ sideband transition in the presence of ancilla $T_1$ decay, the other is limited fidelity (estimated about 85\%) in preparing the ancilla $\ket{f}$ state.  When these two factors are taken into account, the experimental data agrees well with the model prediction (purple).

\section{Outlook}

In this work, we have realized a dispersive type of non-reciprocal interaction between a qubit and a linear cavity in superconducting circuit QED.  This effective interaction, manifested as asymmetric frequency pulls without direct excitation exchange, is mediated by a dissipative bath with broken time reversal symmetry.  The central result of this work is the introduction and verification of a general one-qubit one-cavity master equation model, Eq.~\eqref{eq:master_eq}.   
This model extends the ubiquitous qubit-cavity dispersive interaction in cavity QED to a dissipative setting, and proves to make simple effective predictions of qubit-cavity joint dynamics without tackling the complexity of the bath.  We verify the efficacy of this master equation model through measurements of the qubit dynamics interacting with continuous cavity drive, initial cavity coherent states, and single-photon states.

While the use of non-reciprocity in superconducting circuit QED so far has been primarily limited to canonical circulators in peripheral input/output settings, substantial efforts are under way to integrate non-reciprocal elements with the core part of the quantum devices.  These studies, for example, range from the development of on-chip superconducting parametric circulators for qubit readout~\cite{Sliwa_2015,Chapman_2017,Abdo_2019}, the use of commercial circulators as directional links between quantum modules~\cite{Axline_2018,Kurpiers_2018}, to the development of one-way emitters in waveguide QED~\cite{Kannan_2023,Joshi_2023} 
and the realization of chiral cavity QED for topological many-body physics~\cite{Owens_2022,Roushan_2017}. 
As bath-mediated non-reciprocity becomes more deeply embedded in the devices, complex interactions between quantum modes and components will arise beyond the dichotomy of direct (capacitive/inductive) coupling and cascaded coupling (via a circulator).  To design and to characterize these devices with non-reciprocal elements, it is crucial to have an effective model that are general enough to capture the main features and imperfections of non-reciprocity but also simple enough to not invoke the dynamics of the bath.  Our work provides a first example in this direction.

Beyond circuit QED, demonstration of dispersive non-reciprocity opens a frontier in the study of non-reciprocity beyond standard scattering-type interactions.  For example, making use of the intrinsic connection between those nonreciprocal dynamics and measurement-and-feedforward processes~\cite{Wiseman_1994,Yuxin_2023}, having access to those dynamical elements could enable realization of passively protected quantum memory~\cite{Lieu_2022} and even autonomous quantum error correction~\cite{Xu_2022}. Dispersive nonreciprocity also provides a previously unexplored building block of driven-dissipative quantum many-body systems, and it will be interesting to explore entanglement generation and dissipative phases of quantum matter that arise due to the interplay between Hamiltonian and nonreciprocal dispersive interactions.


\begin{acknowledgements}
We thank Ebru Dogan for experimental assistance. 
This research was supported by the U.S. Department of Energy, Office of Science, National Quantum Information Science Research Centers, Co-Design Center for Quantum Advantage under contract DE-SC0012704.  Initial development of the device package were supported by the Army Research Office Grant No.~W911-NF-17-10469.  Transmon qubits were fabricated and provided by the SQUILL Foundry at MIT Lincoln Laboratory, with funding from the Laboratory for Physical Sciences (LPS) Qubit Collaboratory. A.~A.~C. and Y.-X. W. acknowledge support from the Simons Foundation, through a Simons Investigator award (Grant No.~669487), and 
partial support by the University of Chicago Materials Research Science and Engineering Center, which is funded by the National Science Foundation under award number DMR-2011854.

\end{acknowledgements}

\bibliography{V2}

\clearpage 

\begin{center}
\textbf{\large Supplementary Materials}
\end{center}
\setcounter{equation}{0}
\setcounter{figure}{0}
\setcounter{table}{0}
\setcounter{page}{1}
\setcounter{section}{0}
\makeatletter
\renewcommand{\theequation}{S\arabic{equation}}
\renewcommand{\thefigure}{S\arabic{figure}}

\maketitle

\section{Theory}

\label{subsec:SI.theory}

\subsection{Derivation of the effective master equation with a dissipative dispersive-type interaction}
\label{subsec:adia.elim.multi}

For concreteness, and motivated by the experimental setup discussed in the main text, we consider a multimode system consisting of one cavity mode $ c _{1} $, as well as additional waveguide and circulator modes evolving at timescales much faster than the cavity-qubit system of interest, which are denoted by 
$ c _{j} $ ($ j = 2,3 \ldots , N $). We further introduce a qubit dispersively coupled to the waveguide modes. In the following discussions, we assume the qubit is strongly coupled to a single waveguide mode $ c _{2} $, which corresponds to the bare mode of the buffer cavity for the experimental systems considered in this work. Nevertheless, we note that the results in this section, i.e.~the general form of the effective master equation as well as the generalized Onsager-type reciprocity relations, also hold for the more general case where the qubit is dispersively coupled to multiple intermediary modes. 

The total system dynamics can be described by a Lindblad master equation, which is given by (setting $\hbar=1$ for convenience)
\begin{align}
\label{seq:multi.qme.gen}
\partial_t \hat \rho 
=  & - i \left[ 
	\hat H _{0} + \omega_q \hat\sigma_z
	+ \frac{ \lambda _{0} }{2}  \hat \sigma_z \hat c _{2} ^\dag \hat c _{2} 
	 , \hat \rho   \right]  
	 + \mathcal{L} _{\mathrm{diss}}
	 \hat \rho 
	 ,
\end{align}
where $\hat{H}_0$ describes the Hamiltonian of the couple linear-mode system, and $\mathcal{L} _{\mathrm{diss}}$ encodes the dissipative dynamics of the total cavity-circulator system. Throughout our discussion, we assume the dynamics of the bosonic modes alone is quadratic, so that we have 
\begin{align}
\label{seq:multi.H0.lin}
& \hat H _{0} =  
\sum_{ \ell , m = 1} ^{ N } 
h _{\ell m}  
\hat c _\ell^\dag \hat c _m
,\\
\label{seq:multi.lind.diss}
& \mathcal{L} _{\mathrm{diss}}
    \hat \rho 
    =\sum_{ \ell , m = 1} ^{ n } 
    \Gamma_{\ell m}  
	\left( \hat c _m \rho \hat c _\ell^\dag 
	- \frac{ 1 }{2} \left\{  \hat c _\ell^\dag \hat c _m , 
	\hat \rho  \right \}  \right)   
	. 
\end{align}
Note that for systems with an external magnetic field $B$, the system parameters $ h _{\ell m}  $ and $\Gamma_{\ell m} $ will also depend on the external field $B$. For convenience, we omit this parametric $B$ dependence when it does not cause confusion.

Before proceeding, it is worth noting that the dispersive interaction in Eq.~\eqref{seq:multi.qme.gen} emerges in the deeply detuned limit from a microscopic Jaynes-Cummings coupling between the qubit and mode $c _{2}$, which can be written in the rotating frame at the qubit frequency $\omega_q$ as (ignoring other modes for clarity)
\begin{align}
\label{seq:h.JC}
\hat H _{\mathrm{JC}} 
=  & \Delta _{\mathrm{JC}}
\hat c _{2} ^\dag \hat c _{2} 
+
g \left( 
\hat c _{2} ^\dag \hat \sigma_{-} + 
\hat c _{2} \hat \sigma_{+}  
 \right)   
	 .
\end{align}
Here, $\Delta _{\mathrm{JC}}$ denotes the detuning between qubit and mode $c _{2}$, 
$\Delta _{\mathrm{JC}} \equiv 
\omega_{2} - \omega_{q} $. Thus, our discussions are valid if the following hierarchy in parameters hold:
\begin{align}
\left| \Delta _{\mathrm{JC}} \right| 
\gg g  , \Gamma_{22}  . 
\end{align}
We further assume the system is in the parameter regime where it is valid to adiabatically eliminate the waveguide and circulator modes, i.e.~modes $c _{j} $ with $j>1$. In this case, one can derive an effective Markovian master equation only involving cavity mode $c _{1}$ and the qubit, which can be generally written as 
\begin{align}
\label{seq:qme.eff.1m}
& \partial_t \hat \rho 
	=
	- i [ \hat H _\mathrm{eff}   
	, \hat \rho ] 
	+ \mathcal{L} ^{\mathrm{(eff)}}_{\mathrm{diss}}
    \hat \rho 
    . 
\end{align} 

In what follows, we relabel the cavity mode $c _{1}$ to be the $a $ mode, in accordance with the main text discussion. Noting that the original multimode system dynamics conserves the qubit Pauli operator $\hat \sigma_z $ exactly, and that the dynamics of cavity and circulator modes conditioned on qubit $\hat \sigma_z $ eigenstates are fully quadratic, one can show that the symmetry and dynamical constraints require the effective dynamics to take the following form
\begin{align}
\label{seq:h.eff.1m}
\hat H _\mathrm{eff} &= 
\Delta_c  
\hat a ^\dag \hat a  
	+ \frac{ \lambda  }{2} \hat \sigma_z  \hat a ^\dag \hat a   
, \\
\label{seq:dissp.eff.1m}
\mathcal{L} ^{\mathrm{(eff)}}_{\mathrm{diss}}
& = \kappa  
\mathcal{D} \left[   \hat a \right]   
+ \Gamma   
\mathcal{D} \left[  
e  ^{\frac{ i\theta  + \eta  }{2}
\hat \sigma_z }
\hat a  \right]   
	.
\end{align} 
Here, in the dispersive limit, we have expressed the Hamiltonian in a rotating frame of the qubit and the cavity, where the qubit excitation rotates at the bare qubit frequency $\omega_q$, and the cavity excitations rotate at a fixed reference frequency $\omega_r$ that is close (within several MHz) to the cavity resonance.  Conveniently, $\omega_r$ also corresponds to the frequency of the microwave pulse that we use to initialize cavity coherent states.  We fix $\omega_r$ throughout the experiment across different fields so that the field dependence of the cavity frequency is reflected in the parameter $\Delta_c$.  

For our discussion, it is also convenient to write the quantum Langevin equations for the effective system dynamics, i.e.~ 
\begin{align}
\label{seq:leq.eff.1m}
i \partial_t  \hat a = & 
\mathcal {E} _{ \sigma_z }   \hat a 
- i \sqrt{   \kappa  } \hat  \xi_{\mathrm{in}} 
- i \sqrt{   \Gamma  } 
e  ^{\frac{ -i\theta  + \eta  }{2}
\hat \sigma_z }
\hat  \eta_{ \mathrm{in}} 
	,\\
i \partial_t  \hat \sigma_- 
= &  \Lambda  
	\hat a  ^\dag \hat a  \hat \sigma_- 
+ 2 i \sqrt{   \Gamma  }
 (\sinh \frac{ i\theta  + \eta  }{2})
\hat  \eta_{ \mathrm{in}}  ^\dag 
\hat \sigma_- \hat a  
    \nonumber \\
    & - 2 i \sqrt{ \Gamma   }
(\sinh \frac{ -i\theta  + \eta  }{2})
\hat a ^\dag \hat \sigma_- 
\hat  \eta_{ \mathrm{in}}
	, 
\end{align} 
where $\xi_{\mathrm{in}} $ and $\hat  \eta_{ \mathrm{in}}$ denote standard Markovian quantum input noise operators. To simplify the notation, we also introduce the effective qubit-state-dependent cavity $a $ mode self-energy 
$\mathcal {E} _{ \sigma_z } $, as well as the cavity-to-qubit coupling coefficient 
$\Lambda  $, which can be expressed as 
\begin{align} 
\mathcal {E} _{ \sigma_z } 
= &  \Delta_c  
	 + \frac{ \lambda  }{2}  \sigma_z 
	 - i \frac{ \kappa + \Gamma  
e ^{  \eta \sigma_z  }   }{2} 
\label{seq:eff.a1.energy}
	 , \\
\label{seq:leq.eff.1m.qbcoeff}
\Lambda  
= &  \lambda  
    + i \Gamma ( e ^{ i \theta}  
    - \cosh \eta ) 
	.
\end{align}

Making use of standard adiabatic elimination techniques (see e.g.~\cite{Gardiner_2004book}), we can now extract the effective parameters in terms of zero-frequency susceptibility functions of the original multimode systems (see Eq.~\eqref{seq:multi.H0.scpt}). Before presenting the explicit results, for convenience we introduce the shorthand notation for zero-frequency linear response susceptibilities, as well as the determinant of the zero-frequency susceptibility submatrix of cavity modes, as 
\begin{align}
\chi _{\ell m}   ^{(0)} 
\equiv & \chi _{\ell m}  
\left [ \omega = 0 \right ]
, \\ 
\det \chi ^{(0)} _{\{1,2\} }
\equiv & \chi _{ 11 } ^{(0)} \chi _{ 22 }  ^{(0)} 
	- \chi _{ 12 } ^{(0)}  \chi _{ 21 } ^{(0)}  
.
\end{align}  
We can now relate the effective model parameters to the linear response functions of the original model via the following equations  
\begin{align}
& \mathcal {E} _{ \sigma_z }  
	= - \frac {1 - \frac{ \lambda  _{0} }{2}   
	\chi _{ 22 }  ^{(0)} 
	\sigma_z }
	{\chi _{ 11 } ^{(0)} 
	- \frac{ \lambda  _{0} }{2} \sigma_z 
    \det \chi ^{(0)} _{\{1,2\} }  
	}
	, \\
& \Lambda 
= \frac{ \lambda _{0} 
	\left| \chi _{ 21 } ^{(0)} \right | ^{2}  }
	{ \left\{  \chi _{ 11 } ^{(0)} 
	- \frac{ \lambda  _{0} }{2} 
	\det \chi ^{(0)} _{\{1,2\} }  
	\right\}  
	\left\{  \chi _{ 11 } ^{(0)*} 
	+  \frac{ \lambda  _{0} }{2} 
	\det \chi ^{(0)*} _{\{1,2\} }  
	\right\} } 
	. 
\end{align} 
All individual effective parameters can thus be explicitly derived from above equations; we omit the detailed expressions for compactness.

\subsection{Derivation of the generalized Onsager-type symmetry relations for the dissipative dispersive system}

\label{subsec:Onsager.disp}

In standard linear systems, the microscopic time reversal symmetry (TRS) naturally imposes a constraint on the system's scattering matrix, which is also known as the Onsager-Casimir reciprocity relations~\cite{casimir_1945} in the presence of magnetic fields. Here, we show such reciprocity relations can be generalized  to nonlinear systems with dispersive couplings. 

For reference, let us first review standard Onsager-Casimir reciprocity relations~\cite{casimir_1945}. In terms of linear response susceptibilities $\chi _{\ell m}  
\left [ \omega \right ]
\equiv - i \int ^{+\infty} _{0}
\left \langle  \left[ 
\hat c _{\ell}  \left ( t \right )
, \hat c _{m}^{\dag} \left ( 0 \right ) \right]
\right \rangle 
e ^{i \omega t } dt
$, the Onsager-Casimir relations can be written in the frequency space as  
\begin{align}
\label{seq:recip.lin.multi.gen}
\chi _{\ell m}  
\left [ \omega ; B\right ]
=  \chi _{ m \ell }  
\left [  \omega ; -B\right ] 
.  
\end{align}
For linear systems (e.g.~setting $\lambda = 0 $ in Eq.~\eqref{seq:multi.qme.gen}), the constraints in Eq.~\eqref{seq:recip.lin.multi.gen} is equivalent to the corresponding symmetry conditions on the Hamiltonian and dissipator matrices. To see this, we can consider the Langevin equations of motion of cavity and circulator modes, which is given by 
\begin{align}
\label{seq:multi.H0.qle}
i \partial_t \hat c _\ell  
=  \left(  h _{\ell m}  
- \frac{i}{2}
\Gamma_{\ell m}  
\right)  \hat c _{m} 
+ i  \hat  \xi_{\ell,\mathrm{in}}  
, 
\end{align}
from which we can explicitly write the linear response susceptibilities as 
\begin{align}
\label{seq:multi.H0.scpt}
\chi _{\ell m}  
\left [ \omega \right ]
& = \left [ \left(  \omega
- \mathbf{H} 
+ \frac{i}{2}
\boldsymbol{\Gamma}  
\right) ^{-1} 
\right ]_{\ell m}  
. 
\end{align}
Here, we use boldface letters $\mathbf{H} $ and $\boldsymbol{\Gamma}  $ to denote the Hermitian matrices formed by Hamiltonian (Eq.~\eqref{seq:multi.H0.lin}) and dissipator (Eq.~\eqref{seq:multi.lind.diss}) coefficients, respectively.  
From Eq.~\eqref{seq:multi.H0.scpt}, one can show the constraints in Eq.~\eqref{seq:recip.lin.multi.gen} is equivalent to the following conditions on the Hamiltonian and dissipator matrices, as 
\begin{align} 
\mathbf{H}  \left ( B \right )
& = \left[ \mathbf{H} 
\left ( -B \right )
\right ] ^{T}
, \\  
\boldsymbol{\Gamma}  \left ( B \right )
& = \left[ \boldsymbol{\Gamma}  
\left ( -B \right )
\right ] ^{T}
. 
\end{align}

From Onsager-type reciprocity relations on linear response functions in Eq.~\eqref{seq:recip.lin.multi.gen}, one can show that the effective cavity $a $ mode self-energy 
$\mathcal {E} _{ \sigma_z } $ also needs to satisfy the corresponding reciprocity relation, as 
\begin{align} 
& \mathcal {E} _{  \sigma_z }  \left ( B \right )
= \mathcal {E} _{  \sigma_z } 
\left ( -B \right ) 
. 
\end{align}
From definition of 
$\mathcal {E} _{  \sigma_z } $ in Eq.~\eqref{seq:eff.a1.energy}, we can further show that the following (combinations of) effective quantum master equation parameters must also be symmetric under $B \to -B$:
\begin{align} 
& \Delta_c  
, \, \lambda  
, \, \Gamma  
\sinh  \eta
, \, \kappa  
+ \Gamma  
\cosh \eta
	 . 
\end{align}
We note that the symmetry constraint on  
$\mathcal {E} _{  \sigma_z } $ can be intuitively viewed as a specific $1$-mode case of the more general time reversal symmetry of the dynamical matrix of a linear system. In contrast, the cavity-to-qubit $\Lambda $ is directly related to the squared norm of response function 
$ | \chi _{ 21 } ^{(0)}   | ^{2} 
= \left| \chi _{ 21 } \left [ \omega = 0 \right ] \right | ^{2}$ from cavity modes $1$ to $2$, and hence can generally be asymmetric. As a result, the corresponding effective model parameters $\Gamma \sin\theta $ and $\Gamma \cos\theta $ are also in general asymmetric when flipping the sign of the magnetic field $B$.


\subsection{Solution to the effective quantum master equation with cavity drives}
\label{subsec:simulation_detail}

In the main text, we present comparisons between theoretical predictions of the effective master equation~\eqref{eq:master_eq} (i.e.~Eq.~\eqref{seq:qme.eff.1m}), and experimental measurements of time evolution of the qubit coherence function $\langle 
\hat \sigma_-  (t) \rangle $, due to the application of either a short cavity pulse at the start of the protocol, or continuous drive. Here, we provide a detailed discussion on the techniques used to obtain the theoretical predictions in Figs.~\ref{fig:cw} and~\ref{fig:time} (see solid curves in both figures). For more details on using this general framework to extract master equation parameters and experimental validation based on continuous wave measurement, see Sec.~\ref{subsec:params_extraction} and Sec.~\ref{subsec:validation_cavdr}, respectively.

Our goal is to solve the effective quantum master equation under a cavity linear drive, which can be written in rotating frame for both the qubit and the cavity, with qubit rotates at its bare frequency $\omega _{q}$, cavity and cavity linear drive rotates at the reference frequency $\omega_r$ as 
\begin{align} 
\partial_t \hat \rho 
=   & -i \left[ \Delta_c
\hat a ^\dag \hat a   
+ \frac{ \lambda  }{2} \hat \sigma_z  \hat a ^\dag \hat a + \hat H_{\mathrm{dr}}, \hat \rho   \right]
+ \mathcal{L} ^{\mathrm{(eff)}}_{\mathrm{diss}}  
\hat \rho   
, 
\end{align} 
where the linear drive on the cavity is described by the Hamiltonian
$\hat H_{\mathrm{dr}} = 
\epsilon (t) (\hat{a}^\dagger e^{-i\Delta_{ \mathrm{d} } t} + h.c.) $, the dissipative term 
$\mathcal{L} ^{\mathrm{(eff)}}_{\mathrm{diss}} $ is defined in Eq.~\eqref{seq:dissp.eff.1m} (see also Eq.~\eqref{eq:master_eq}), and 
$\Delta_{c} $ denotes detuning between cavity mode $a$ frequency and reference frequency, $\Delta_{c} = \omega_c - \omega_r$, and $\Delta_{d} $ denotes detuning between drive frequency and reference frequency, $\Delta_{d} = \omega_{ \mathrm{dr} } - \omega_r$. Note that for the qubit Ramsey experiments with a pulsed cavity drive, the cavity drive is set to be resonant with $\Delta_{d} = 0$. 
While the following discussion is applicable to pulsed and continuous drives alike, to simplify notations, we abbreviate the drive amplitude $\epsilon (t) $ as $\epsilon$ when it does not cause confusion. We first go into the cavity rotating frame with respect to drive detuning $\Delta_{ \mathrm{d} }$, so that we have   
\begin{align} 
\partial_t \hat \rho 
=   & -i \left[ 
(\Delta_c - \Delta _{ d } )\hat a ^\dag \hat a  
+ \frac{ \lambda  }{2} \hat \sigma_z  \hat a ^\dag \hat a + \epsilon (\hat{a}^\dagger  + \hat{a}) , 
\hat \rho   \right] 
\nonumber \\
& + \mathcal{L} ^{\mathrm{(eff)}}_{\mathrm{diss}}  
\hat \rho    
, 
\end{align} 
The qubit coherence function 
$\langle \hat \sigma_-  (t) 
\rangle   $ 
can thus be computed by solving the following equation of motion 
 ($\left \langle \uparrow \right |
\hat \rho  
\left | \downarrow\right \rangle 
\equiv \hat \rho _{\uparrow \downarrow} $)
\begin{align} 
\label{seq:qme.eff.1mdr.coh}
\partial_t \hat \rho _{\uparrow \downarrow} 
=   & -i \left[ 
(\Delta_c - \Delta _{ d } ) \hat a ^\dag \hat a  
+ \epsilon (\hat{a}^\dagger  + \hat{a}) , 
\hat \rho _{\uparrow \downarrow}  \right]
\nonumber \\
& - i \frac{ \lambda  }{2} \left \{
\hat a ^\dag \hat a 
, \hat \rho _{\uparrow \downarrow} \right \}
+ \mathcal{L} ^{\mathrm{(eff)}}_{\mathrm{diss}}  
\hat \rho _{\uparrow \downarrow} 
. 
\end{align}

Before we proceed, it is worth noting that, in principle, one could try to simulate Eq.~\eqref{seq:qme.eff.1mdr.coh} with a brute-force approach, by directly numerically evolving the quantum master equation in the joint basis between qubit states 
$\left | \uparrow \right \rangle $, 
$\left | \downarrow \right \rangle $ and cavity photon Fock states. In this case, the simulation cost would depend on the transient cavity photon number, and grows as the cavity pump power is increased. Here, we take a approach whose computation cost does not scale with photon number, making use of the fact that Eq.~\eqref{seq:qme.eff.1mdr.coh} describes a quadratic superoperator on the coherence operator 
$\hat \rho _{\uparrow \downarrow} $, so that Gaussian states stay Gaussian. As result, the dynamics in Eq.~\eqref{seq:qme.eff.1mdr.coh} could be solved making use of standard phase-space-based methods~\cite{Gardiner_2004book}. However, noting that the dissipative dynamics in Eq.~\eqref{seq:dissp.eff.1m} do not involve any heating dissipator, this procedure can be further simplified, by taking the ansatz 
\begin{align} 
\label{seq:1m.coh.pure}
\hat \rho _{\uparrow \downarrow} (t)
= \left | \bar{a} _{\uparrow}  (t) \right \rangle 
\left \langle \bar{a} _{\downarrow}  (t) 
\right |
. 
\end{align} 
Substituting Eq.~\eqref{seq:1m.coh.pure} into Eq.~\eqref{seq:qme.eff.1mdr.coh}, one can show that when the conditional amplitudes satisfy the following linear equations of motion, Eq.~\eqref{seq:qme.eff.1mdr.coh} is also satisfied
\begin{align} 
\label{seq:eom.eff.1m.avg}
i \partial_t  \bar{a} _{\sigma_z}   = & 
\mathcal {E} _{ \sigma_z }   
\bar{a} _{\sigma_z} 
+ \epsilon
,  \quad 
( \sigma_z = \uparrow , \downarrow)
, 
\end{align} 
where the effective energies of the conditional cavity amplitudes can again be calculated using Eq.~\eqref{seq:eff.a1.energy}, but with the drive detuning $\Delta_c - \Delta _{ d } $. Note that the first term in the right hand side of Eq.~\eqref{seq:eom.eff.1m.avg} can be intuitively viewed as taking the expectation value of the Langevin equation~\eqref{seq:leq.eff.1m}, ignoring the noise terms. In this case, we can also calculate the qubit coherence function via the equation below
\begin{align} 
\label{seq:eom.eff.qbcoh}
&   \frac{ d \langle 
\hat \sigma_-  (t) \rangle }
{  dt } 
= \Lambda \bar{a} _{\uparrow} 
\bar{a} ^*_{\downarrow} 
\langle 
\hat \sigma_-  (t) \rangle
\nonumber \\
= & \left[ -  i \lambda  
    +\Gamma  (e ^{ i \theta}  
    - \cosh \eta ) 
    \right] 
    \bar{a} _{\uparrow} 
    \bar{a} ^*_{\downarrow} 
    \langle 
\hat \sigma_-  (t) \rangle
, 
\end{align} 
where the coefficient $\Lambda $ is given in Eq.~\eqref{seq:leq.eff.1m.qbcoeff}. The above equation~\eqref{seq:eom.eff.qbcoh} reproduces Eq.~\eqref{eq:eom_qb_coh} in the main text. Note that we have reduced the problem of simulating the equation of motion~\eqref{seq:qme.eff.1mdr.coh} of an operator acting on the cavity Hilbert space, down to integrating the set of linear differential equations in Eq.~\eqref{seq:eom.eff.1m.avg}, which greatly improves the efficiency of the numerical calculations.


\subsection{Master equation parameter extraction}
\label{subsec:params_extraction}
The quantum master equation~\eqref{eq:master_eq} in the main text contains 6 independent real parameters. To extract those variables for a given experimental setup, we need equivalent amount of measurements. More specifically, one can show that the master equation parameters can be uniquely determined by measuring (1) cavity frequencies and lifetimes contingent on qubit states, and (2) cavity-induced qubit phase shift and dephasing factor. To see this, we first note that the cavity resonances $ \omega _{g, e}$ and linewidths $\kappa_{g, e}$ 
let us obtain the following relations about system parameters: 
\begin{align}
\Delta_c = &  
(\omega_{g} + \omega_{e} )/2 - \omega_r , 
\\
\lambda =& \omega_{g} - \omega_{e} , 
\label{eq:lambda_eff}
\\
\Gamma \sinh \eta =& \kappa_{g}-\kappa_{e} , 
\label{eq:Gamma_Re_r}
\\
\kappa + \Gamma \cosh \eta = & (\kappa_{g}+\kappa_{e})/2 
. 
\label{eq:kappa_eff}
\end{align}

For the qubit measurements, integrating Eq.~\eqref{eq:eom_qb_coh} (i.e.~Eq.~\eqref{seq:eom.eff.qbcoh}) over time leads to the following equaton relating the qubit phase shift and decoherence factor to system parameters, as 
\begin{align}
\ln \frac{ \left\langle\hat{\sigma}_{-}\left(t_{f}\right)\right\rangle}{\left\langle\hat{\sigma}_{-}(0)\right\rangle} 
&\equiv  
(i \phi + \ln\zeta) 
\label{eq:coherence_evolution}
\\
\nonumber
&= \bar{a} _{\uparrow} (0)
\bar{a} ^*_{\downarrow} (0)
\frac{
-i\lambda +  \Gamma (e ^{ i \theta}  
- \cosh \eta ) }
{i\lambda +\kappa +\Gamma \cosh \eta 
}\\ \nonumber
&\times \left(1-
e^{-\left ( i\lambda +\kappa 
+\Gamma \cosh \eta \right) t_{f} }\right)
. 
\end{align}
where $\bar{a} _{\uparrow/\downarrow} (0)$ are the initial cavity amplitudes conditioning on qubit state. Assuming that $ \bar{a} _{\uparrow}(0) 
\approx \bar{a} _{\downarrow} (0)$ (short drive pulse limit), we can factor out the drive dependent term 
$\bar{a} _{\uparrow} (0)
\bar{a} ^*_{\downarrow} (0)$ by rescaling the equation with respect to the cavity photon number at $t = 0$, $\bar{n}_0$. We can thus obtain the following equations connecting parameters to qubit measurements 
\begin{align}
&\Gamma \sin \theta = \lambda
 + \mathrm{Im} \left [
\frac{i\lambda  + (\kappa_{g}+\kappa_{e})/2}{1-e^{(-i\lambda-(\kappa_{g}+\kappa_{e})/2)t_f}} 
\frac{i \phi + \ln\zeta}{ \bar{n}_0 } \right]
, \label{eq:Gamma_Im_r}
\\
&\Gamma (\cosh \eta -\cos\theta)
= -\frac{1}{2} \mathrm{Re}
\Bigg[\frac{i\lambda + (\kappa_{g}+\kappa_{e})/2}
{1- e^{-\left ( i\lambda +\kappa 
+\Gamma \cosh \eta \right) t_{f} }}  
\nonumber \\
& \quad \quad \quad \quad \quad 
\times 
\frac{i \phi + \ln\zeta}{ \bar{n}_0 }
\Bigg]
. \label{eq:Gamma_r_2}
\end{align}
Taking the long evolution time limit, we obtain Eq.~\eqref{eq:qb.ramsey.tf} in the main text.

\subsection{Simulation of the effective master equation with a continuous or pulsed cavity drive}
\label{subsec:validation_cavdr}

From Eqs.~\eqref{seq:1m.coh.pure} and~\eqref{seq:eom.eff.qbcoh}, the system dynamics under continuous drive can be straightforwardly solved. In this case, the drive amplitude $\epsilon$ in Eq.~\eqref{seq:eom.eff.1m.avg} is given by a constant, so that in the stationary state regime, the qubit-state-conditional cavity amplitudes can be solved by setting the right hand sides of Eq.~\eqref{seq:eom.eff.1m.avg} to zero. Substituting these steady-state amplitudes into Eq.~\eqref{seq:eom.eff.qbcoh},  we can analytically derive the qubit Stark shift ($\Delta_q$) and dephasing rate ($\Gamma_q$) as 
\begin{align}
i \Delta_q - \Gamma_q
= \frac{ -  i \lambda  
    +\Gamma  (e ^{ i \theta}  
    - \cosh \eta ) }
{ \mathcal {E} _{ \uparrow } 
\mathcal {E} ^*_{ \downarrow } }
\left | \epsilon ^2 \right |
. 
\end{align}
Invoking this equation, and noting that the conditional cavity frequencies now explicitly depend on the drive detuning 
$\Delta _{ d } $, we can compute the theory curves in Fig.~\ref{fig:cw}.

To produce the theoretical prediction (solid curves) for the case with pulsed cavity drive in Fig.~\ref{fig:time}, we implement a square cavity pulse of width $18~\text{ns}$ in Eq.~\eqref{seq:eom.eff.1m.avg} before the beginning of the qubit measurement sequence, and then evolve Eq.~\eqref{seq:eom.eff.1m.avg} without driving (i.e.~setting $\epsilon(t)=0$) for the duration of the measurement. We then use the numerically solved conditional cavity amplitudes and Eq.~\eqref{seq:eom.eff.qbcoh} to compute the time evolution of qubit coherence function. When comparing the theory predictions with the experimental values, we introduce an additional shift in the zero of the temporal axis, to account for the relatively longer qubit $\pi/2$ pulse during the Ramsey sequence.

\subsection{Simulation of the effective master equation with an initial Fock state, or quasi-single photon dynamics}
\label{subsec:Fock_FWM_Fock}
Here we discuss the numerical methods used to obtain the theoretical prediction in Fig.~\eqref{fig:fwm}, i.e.~simulating Eq.~\eqref{eq:master_eq} either from a single-photon Fock state or from a qubit excited state with an additional four-wave mixing (FWM) pulse. 
For the first case, the qubit coherence function 
$\langle \hat \sigma_-  (t) 
\rangle   $ 
can similarly be computed by solving the following equation of motion
\begin{align} 
\label{seq:qme.eff.1m.coh}
\partial_t \hat \rho _{\uparrow \downarrow} 
=   & -i \left[ 
- \Delta_c \hat a ^\dag \hat a  , 
\hat \rho _{\uparrow \downarrow}  \right]
- i \frac{ \lambda  }{2} \left \{
\hat a ^\dag \hat a 
, \hat \rho _{\uparrow \downarrow} \right \}
+ \mathcal{L} ^{\mathrm{(eff)}}_{\mathrm{diss}}  
\hat \rho _{\uparrow \downarrow} 
. 
\end{align} 
With the cavity starting from a single-photon Fock state as the initial state, the qubit coherence function in the long-time steady state limit can be analytically derived as 
\begin{align} 
\frac{ \left\langle\hat{\sigma}_{-}
\left( \infty \right)
\right\rangle}
{\left\langle\hat{\sigma}_{-}(0)
\right\rangle} 
=&  \frac{
\kappa + \Gamma e ^{ i \theta}   }
{i\lambda +\kappa +\Gamma \cosh \eta 
}
, 
\end{align} 
which lets us compute the yellow squares in Fig.~\ref{fig:fwm}.

For the FWM case, the qubit is initialized into a state with nonzero population in the higher-level 
$\ket{f}$ state, and the total system then evolves in the presence of the FWM drive that enables the 
$\ket{f0} \leftrightarrow 
\ket{g1}$ transition. In the rotating frame with respect to the FWM drive frequency, the total system master equation can thus be written as 
\begin{align}
\partial_t \hat \rho 
=   & -i \left[ 
\hat H _\mathrm{eff}
+ \Omega  ( 
\hat a ^\dag 
\left | g \rangle
\langle f \right |
+ \text{H.c.} ) , 
\hat \rho   \right]
+ \mathcal{L} ^{\mathrm{(eff)}}_{\mathrm{diss}}  
\hat \rho    
\nonumber \\
\label{seq:eff.qme.fwm}
& + \gamma _{f}  
\mathcal{D} \left[ 
\left | e \rangle
\langle f \right |\right] 
\hat \rho    
+ \gamma _{e}    
\mathcal{D} \left[  
\left | g \rangle
\langle e \right | \right]   
\hat \rho    
.
\end{align} 
where $\hat H _\mathrm{eff}$ and 
$\mathcal{L} ^{\mathrm{(eff)}}_{\mathrm{diss}} $ are again the effective Hamiltonian and dissipator after adiabatic elimination of the other waveguide and circulator modes (see Eqs.~\eqref{seq:h.eff.1m} and~\eqref{seq:dissp.eff.1m}), and $\Omega$ denotes the FWM drive Rabi frequency.
To accurately describe time evolution in this case, we also need to include the $T_1$ decay processes of the excited $\ket{f}$ and $\ket{e}$ states, with corresponding decay rates given by 
$\gamma _{f} $ and $\gamma _{e} $ respectively.

In the experiments, the cavity mode $a$ starts in the vacuum state, whereas the qubit starts in a mixture between 
$\ket{f}$ and $\ket{g}$ states due to initialization imperfections. In this case, time evolution according to Eq.~\eqref{seq:eff.qme.fwm} only involves a small finite number of system levels, so that its numerical integration is straightforward. Making use of the numerical solution to Eq.~\eqref{seq:eff.qme.fwm}, we obtain the theory predictions for an imperfect initial $\ket{f}$ state (purple diamonds) in Fig.~\ref{fig:fwm}.


\section{Experimental setup and techniques}

\begin{figure*}[!htbp]
    \centering
    \includegraphics[scale=.45]{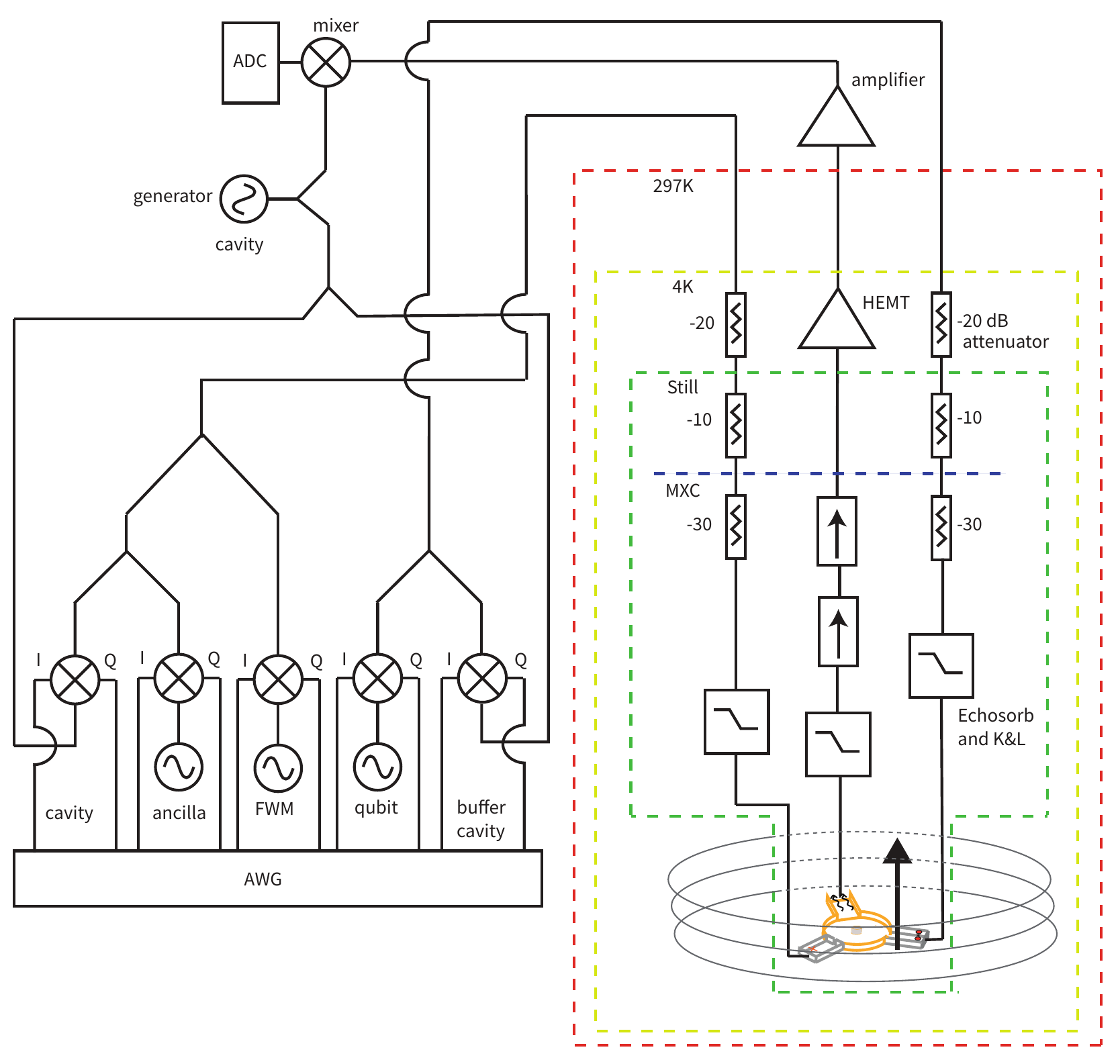}
    \caption{\textbf{Fridge wiring schematic.} Our experiment is performed in a Bluefors LD250 dilution refrigerator. The device is mounted under a mezzanine plate which is thermalized to the MXC plate whose base temprature is around 10 mK. The device is positioned at the center of a superconducting solenoid magnet which operates at 4 K. The qubit port and cavity port of the device are connected to input cables (with attenuators as marked) and the open waveguide port is connected to an output amplifier line.}
    \label{fig:fridge_wiring}
\end{figure*}


\subsection{Device and measurement setup}

The device is mounted under a mezzanine plate which is thermalized to the MXC plate of a Bluefors LD250 dilution refrigerator, as shown in Fig.~\ref{fig:fridge_wiring}, and is positioned at the center of a superconducting solenoid magnet which operates at 4~K. The qubit port and cavity port of the device are connected to input cables (with attenuators as marked) with eccosorb filters and 
a 12 GHz K$\&$L low pass filter to filter high frequency niose and protect the qubit. The open waveguide port is connected to an output amplifier line, so that we can get the cavity transmission measurement results through the output port. The cavity port is used to drive both the cavity and its ancilla qubit which is dispersively coupled to the cavity. The cavity state information is reflected through the read out ancilla qubit state. The qubit port receives a drive for both the qubit and its buffer cavity, with the buffer cavity's transmission signal collected from the same output port. The qubit is dispersively coupled to the buffer cavity, allowing it to act as a proxy for information about the cavity state. The FWM drive, which is for the quasi-single photon measurements, is sent from a separate generator to the cavity port as well.

\begin{figure}[!htbp]
    \centering
    \includegraphics[scale=.5]{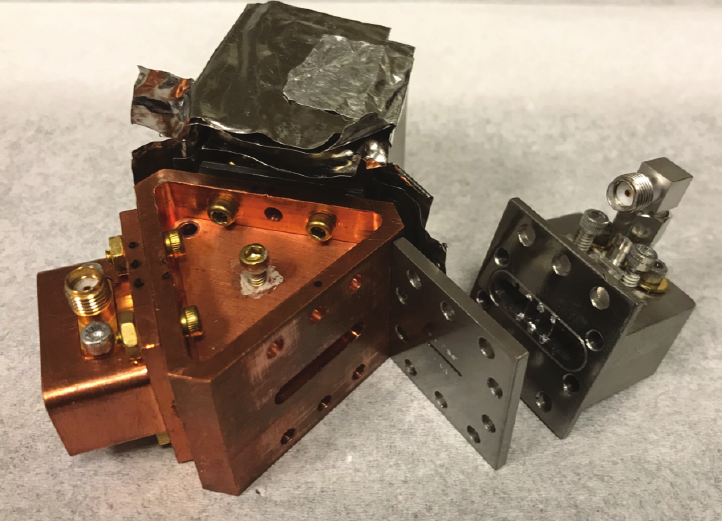}
    \caption{\textbf{Device picture.} Two niobium segments are hooked to a copper intermediary region via niobium plates with apertures to illicit an evanescent coupling. A 21.5~mm $\times$ 5.5~mm sapphire chip hosting a qubit is inserted into the niobium segment slot, with extra indium added at the end of the sapphire slot to ensure the chip is held in place. Two extra blank chips are inserted to maintain the cavity frequency and the symmetry of the electromagnetic field to ensure enough coupling to the intermediary mode. Two tuning screws at the top of the niobium cavity allow fine tuning of the cavity frequency. Both niobium cavities are shielded with extra layer of mu metal to provide further magnetic protection for the qubits.}
    \label{fig:device_picture}
\end{figure}

\begin{figure}[!htbp]
    \centering
    \includegraphics[scale=.24]{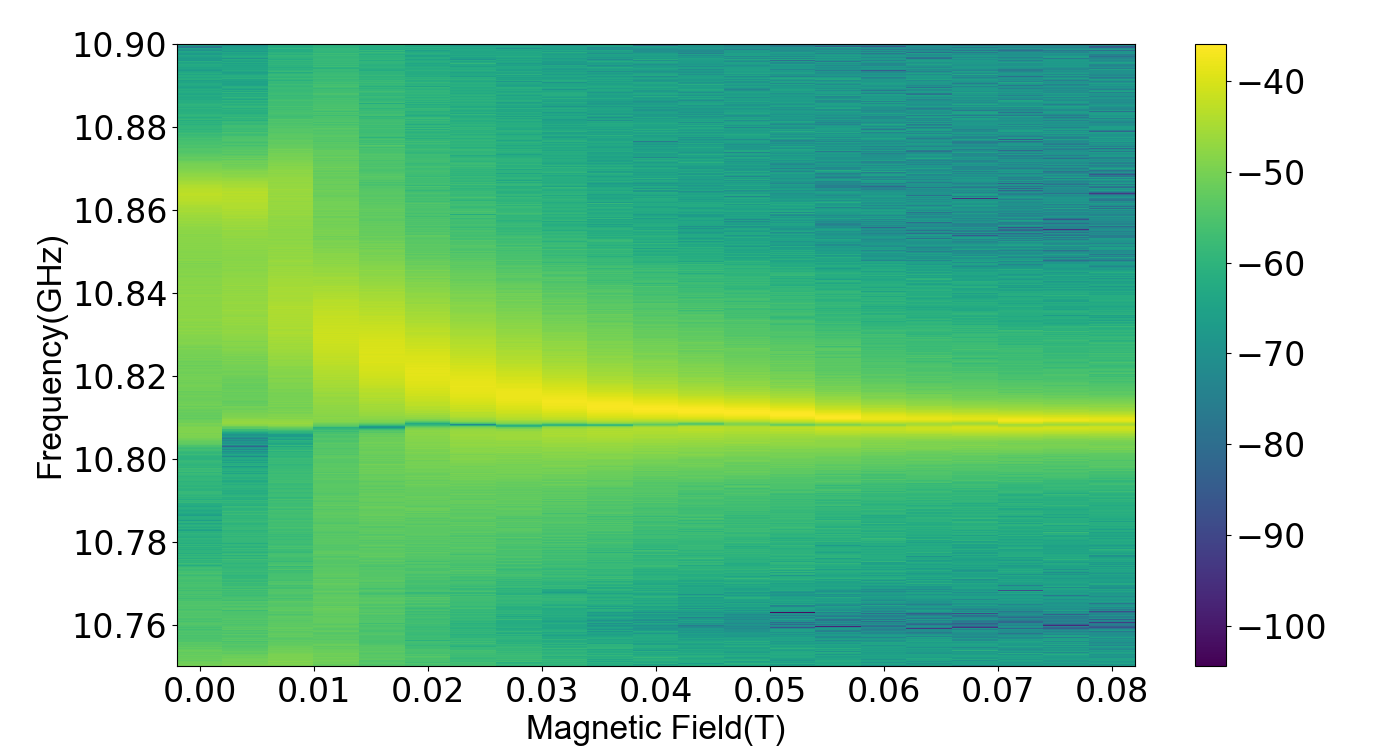}
    \caption{\textbf{Device transmission measurements under various external fields.} VNA transmission measurements, from the buffer cavity port to the output port, under various external magnetic fields. The dominant peak at each field is the buffer cavity mode which is broadband at low field due to the strong coupling to the output port. The dip at around 10.81GHz corresponds to the cavity mode.}
    \label{fig:cavity_spec}
\end{figure}

Both cavities are made out of niobium which is superconducting at low temprature and remains superconducting under the magnetic field applied for this experiment. These cavities are shielded with mumetal sheets to further protect the qubits from the applied magnetic field, as shown in Fig.~\ref{fig:device_picture}. The mumetal sheets are mounted to the cavity for proper thermalization. 

Qubits are integrated with the setup to perform this experiment, as shown in Fig.~\ref{fig:device_picture}, by using a 21.5 mm $\times$ 5.5 mm sapphire chip. This chip hosts a qubit and is inserted into the niobium segment slot, with extra indium added at the end of the sapphire slot to ensure the chip is mechanically held in place and properly thermalized.
Note that there are three sapphire chips in the niobium segment, and the qubit is located at the rightmost chip. The rest of the chips are there for maintaining the symmetry of the electromagnetic field of the cavity mode inside and tuning the cavity mode frequency. To further tune the frequency, two tunning screws at the top of the niobium cavity are used to fine tune the cavity frequency, thus obtaining a cavity frequency of 10.808 GHz and the qubit's buffer cavity frequency 10.809 GHz. As mentioned in the main text, the niobium segment housing the qubit we are interested in is connected to the center region through a plate with 8~mm $\times$ 1~mm aperture, which makes this niobium segment a very broadband cavity. It has a larger than 5 MHz linewidth at all tested fields with a linewidth above 20 MHz at lower fields. The operating cavity uses a 2.5 mm radius circular aperture, making it a relatively low linewidth.

The transmission properties of the device are obtained through a Vector Network Analyzer (VNA), with the VNA ports connected to the buffer cavity port and the output port. Sweeping the fridge magnet, we obtain the spectrum in Fig.~\ref{fig:cavity_spec}.


\subsection{Ramsey measurement protocol and analysis}
\label{subsec:Ramsey}
\begin{figure}[tbp]
    \centering
    \includegraphics[scale=.5]{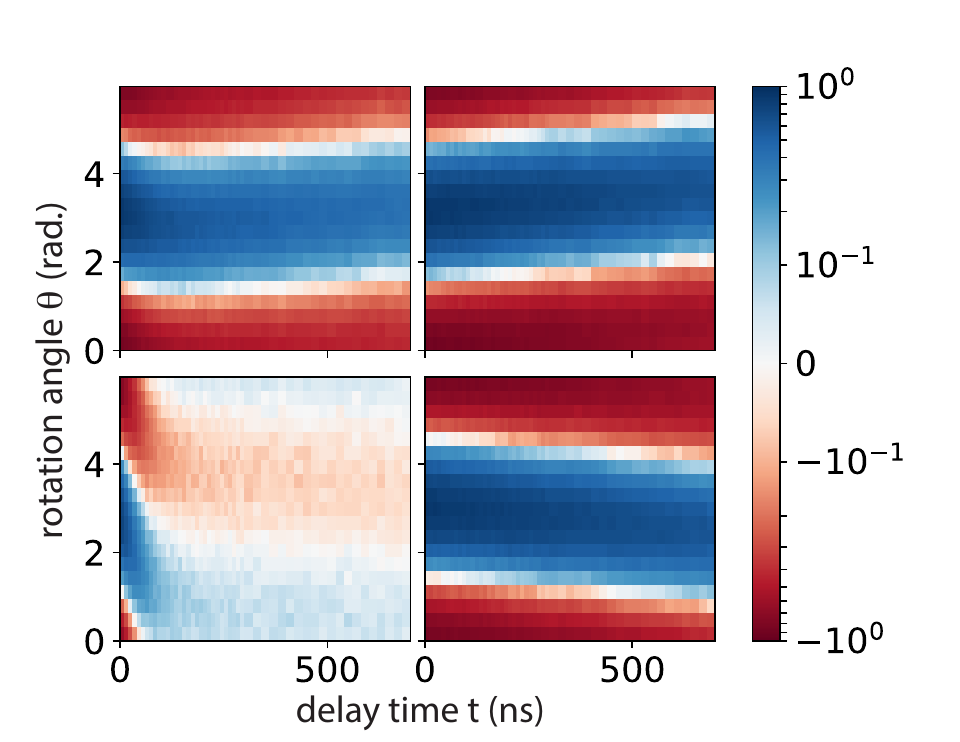}
    \caption{\textbf{Qubit Ramsey measurement results at $\pm 30$ mT.} The raw result of qubit Ramsey measurements shows the qubit state $\sigma_z$ under different delay times as the x-axis ($t$ in Fig.~\ref{fig:ramseys}a). The y axis is the second qubit pulse's rotation angle ($\theta$ in Fig.~\ref{fig:ramseys}a). The upper panels depict the data taken at 30 mT and lower panels taken at $-30$ mT. The left panels show evolution under cavity photons and right panels show just the qubit natural evolution. }
    \label{fig:raw_ramseys}
\end{figure}

Using the measurement protocol as shown in Fig.~\ref{fig:ramseys}(a), the full qubit Ramsey results at 30 mT and $-30$ mT are obtained and shown in Fig.~\ref{fig:raw_ramseys}. We can use this data to extract the qubit phase shift and decoherence due to the cavity photons, 
where the measured phase shift is dominated by the term $\Gamma \sin\theta$ of the master equation as per Eq.~\eqref{eq:Gamma_Im_r}. The decoherence rate is dominated by the term $\Gamma (\cosh \eta -\cos\theta)$, see Eq.~\eqref{eq:Gamma_r_2}. 

\begin{figure}[tbp]
    \centering
    \includegraphics[scale=.5]{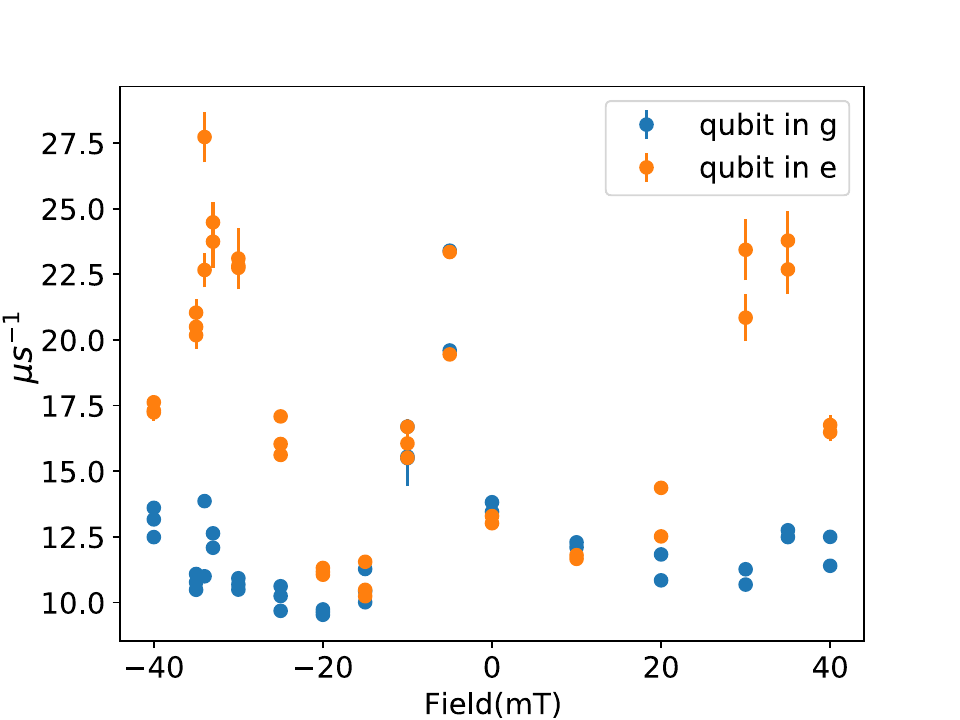}
    \caption{\textbf{Cavity decay rate under varying magnetic field.} The cavity decay rate results based on the same measurement protocol and analysis as shown in Fig.~\ref{fig:cavT1s} across varying magnetic fields.}
    \label{fig:cavity_linewidth}
\end{figure}

Cavity Ramsey measurements, using the measurement protocol in Fig.~\ref{fig:ramseys}(c), provide oscillatory data that is fit with an exponentially decayed sinusoid. The fitted oscillation frequency thus determines the cavity frequency. Note that we do not obtain the cavity decay rate based on the fit exponential decay rate, as the y-axis of the oscillation is directly related to the ancilla state. While this oscillation monotonically depends on the photon number, it is not linearly proportional.  
The cavity decay rate measurement follows the pulse diagram in Fig.~\ref{fig:cavT1s}(a). The cavity decay rates, i.e.~cavity $T_1$ information, are obtained and plotted in Fig.~\ref{fig:cavity_linewidth}. These values can then be used, along with the other measured values, to determine the master equation parameters as laid out in Sec.~\ref{subsec:params_extraction} and shown in Fig.~\ref{fig:params}.

\subsection{Calibration of cavity photon number}
\label{subsec:photon_number_cal}
\begin{figure}[!htbp]
    \centering
    \includegraphics[scale=0.75]{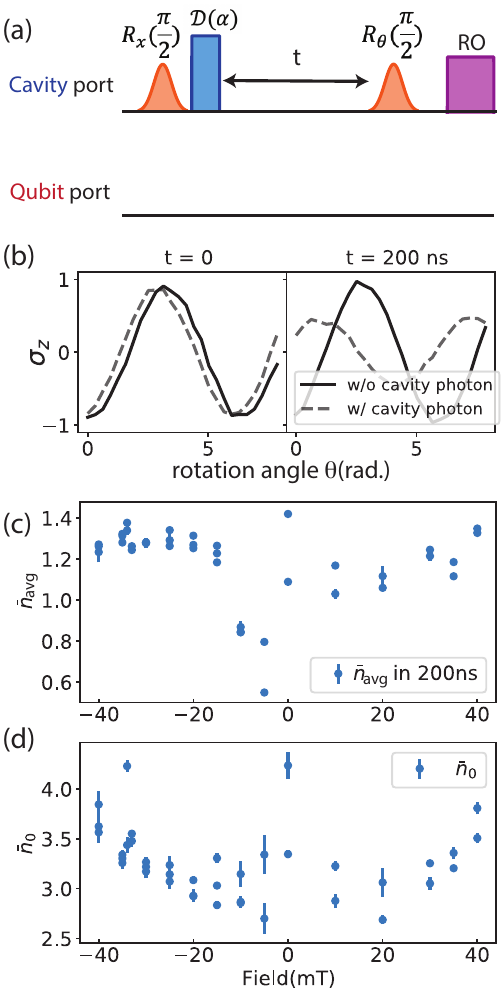}
    \caption{\textbf{Photon number calibration.} (a) The pulse diagram and (b) the example experimental result of cavity photon number calibration for an ancilla qubit Ramsey measurement under cavity photons. 
    We can obtain the ancilla qubit phase evolution by varying the delay time between the cavity pulse and the second ancilla qubit $\pi$/2 pulse. The result of the ancilla qubit state against the rotation angle $\theta$ is plotted in Fig.~\ref{fig:ramseys}(b). We can obtain the phase and the amplitude from these sinusoidal curves.
    The solid and dashed lines are for the results without or with cavity photons respectively, where the phase difference between them is the accumulated phase of qubit due to cavity photons. (c) The averaged cavity photon number during 200 ns, based on the time-averaged frequency shift divided by the dispersive shift $\chi$ between the cavity and the ancilla qubit. (d) Initial photon number calculated with Eq.~\eqref{eq:n_initial}.}
    \label{fig:photon_number}
\end{figure}
The averaged photon number $\bar{n}_\mathrm{avg}$ is required for calculating the qubit frequency shift per cavity photon, $\chi_\mathrm{cq} = \phi/(t\cdot\bar{n}_\mathrm{avg}) $. To calibrate $\bar{n}_\mathrm{avg}$, we measure an ancilla qubit that is directly dispersively coupled to the cavity, where we perform Ramsey experiments on the ancilla with and without cavity photons. By comparing these measurements, we can extract the accumulated extra ancilla phase shift $\phi_{a}$ caused by cavity photons over a finite time window of $t = 200$ ns, as shown in Fig.~\ref{fig:photon_number}(a).
Example results of the ancilla state versus rotation angle $\theta$ of the second Ramsey $\pi/2$ pulse is plotted in Fig.~\ref{fig:photon_number}(b),
with the solid and dashed lines representing cases without and with cavity photons, respectively. 
As $\phi_\mathrm{a}$ is proportional to the $\bar{n}_\mathrm{avg}$, with the ratio set by dispersive coupling strength $\chi_\mathrm{a}$ between the ancilla qubit and the cavity, 
$\phi_\mathrm{a} = \chi_\mathrm{a} \bar{n}_\mathrm{avg} t$,
we obtain $\bar{n}_\mathrm{avg}$ in 200 ns,
\begin{equation}
\bar{n}_\mathrm{avg} = \phi_\mathrm{a} /( \chi_\mathrm{a}\cdot  t),
\label{eq:n_avg}
\end{equation}
and repeat this protocol for all of the external fields, the result is plotted in  Fig.~\ref{fig:photon_number}(c). 

The initial photon number $\bar{n} _0$ is a required quantity for computing the master equation parameters. We can compute its value via the measured cavity decay rate conditioned on the qubit in ground state, $\kappa_{g}$, as well as the time-averaged cavity photon number $\bar{n}_\mathrm{avg}$ during the time window $t$ (see paragraph above for discussion on its measurement), making use of the following equation:
\begin{equation}
\bar{n}_0 = \bar{n}_\mathrm{avg}  \frac{\kappa_{g}t}{1 - e^{-\kappa_{g}t}} . 
\label{eq:n_initial}
\end{equation}
The initial photon number $\bar{n} _0$ is plotted against varying magnetic fields in Fig.~\ref{fig:photon_number}(d).

\subsection{Master equation parameter uncertainty extraction}
\label{subsec:params}
\begin{figure}[tbp]
    \centering
    \includegraphics[width=8.5cm]{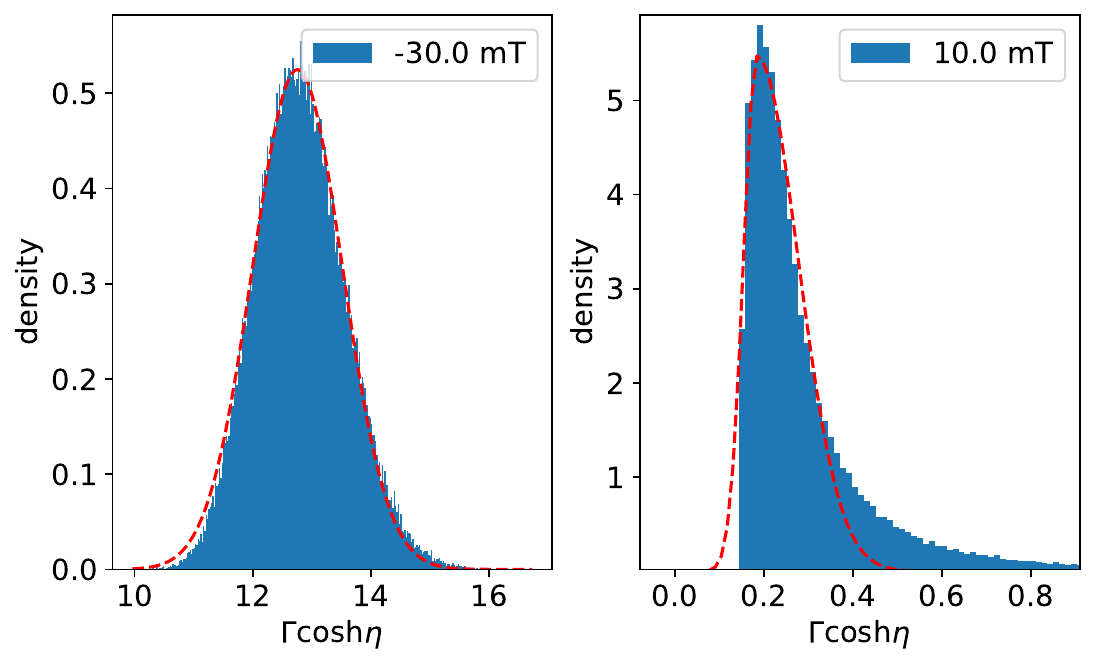}
    \caption{\textbf{Gaussian fits to obtain uncertainty of master equation parameters.} Master equation parameter distributions utilizing a Monte Carlo simulation for the uncertainty extraction for (a) $\Gamma\cosh\eta$ at -30 mT yields a fairly good Gaussian distribution; the Gaussian fit curve (red dashed line) yields $\sigma = 0.76$ as its uncertainty. The extraction for (b) $\Gamma\cosh\eta$ at 10~mT, a noticeably asymmetry distribution; the combination of two half Gaussian fit curves (red dashed line) yields $\sigma = 0.03$ from left half curve as lower bound uncertainty and $\sigma = 0.09$ from right half curve as higher bound uncertainty. The results are derived with the assumption of a Gaussian distribution and uncertainty of each measurement outputs as $\sigma$ of the Gaussian distribution to generate random numbers as an input. $10^5$ simulations are used to extract the output value of the master equation parameters 
    based on the calculation formulae in Sec.~\ref{subsec:params_extraction}. }
    \label{fig:histogram}
\end{figure}

\begin{figure}[tbp]
    \centering
    \includegraphics[width=8.5cm]{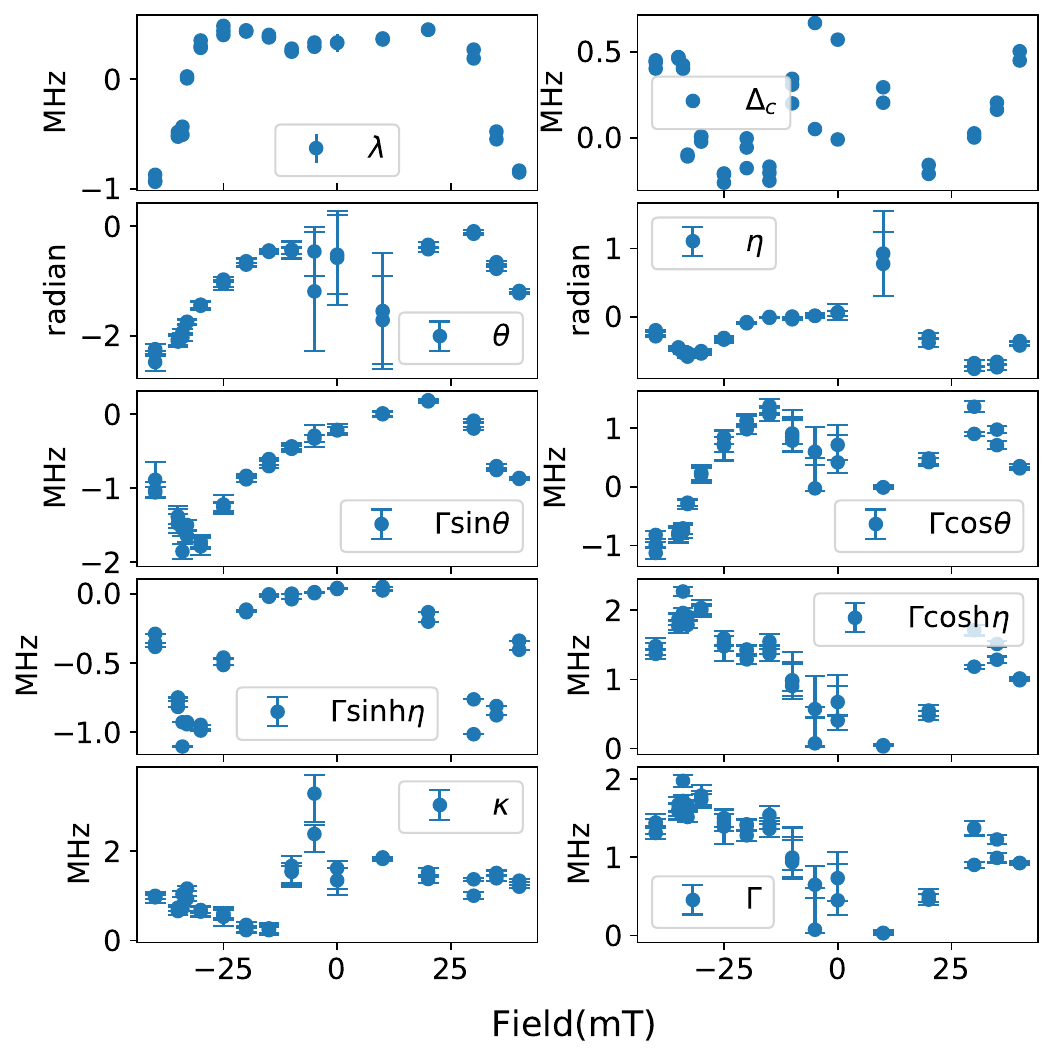}
    \caption{\textbf{Master equation parameters.} The derived master equation parameter values and uncertainties. Each set of the measurement data is used with the Monte Carlo simulation to determine the uncertainties.} 
    \label{fig:all_points}
\end{figure}

The master equation parameter extraction is relatively complex as shown in Sec.~\ref{subsec:params_extraction}, which makes the direct calculation of the parameter uncertainty fairly challenging. Thus, a Monte Carlo method is applied for the uncertainty extraction. We assume each measurement output satisfies a Gaussian distribution, with the standard deviation $\sigma$ of the Gaussian distribution set by the corresponding measurement uncertainty, and perform $10^{5}$ simulation runs with randomly generated input to accumulate the output value statistics of the master equation parameters based on the calculation formulae in Sec.~\ref{subsec:params_extraction}.
For cases shown in Fig.~\ref{fig:histogram}(a), where the master equation parameter generated fits well to a Gaussian distribution, the standard deviation $\sigma$ of the Gaussian curve fit is used as the uncertainty for this parameter. For cases where the distribution shows asymmetry like Fig.~\ref{fig:histogram}(b), where a simple Gaussian distribution cannot provide an accurate fit, we did separate half Gaussian fits for the left and right to obtain the uncertainty of the upper and lower bounds of the parameter. The fitting result of the Gaussian center would be the value obtained for the master equation parameter value for each set of the measurements, as plotted in Fig.~\ref{fig:all_points}.


Due to repeated measurements at the same magnetic field being performed, the result in Fig.~\ref{fig:all_points} has multiple data points. We can then average these points to get a final value and uncertainty at a specific field. There is an intrinsic data point uncertainty and an uncertainty associated with the distribution of the data points. We select the larger of these two values to use as the uncertainty.


\label{subsec:cw}
\begin{figure}[tbp]
    \centering
    \includegraphics[width=8.5cm]{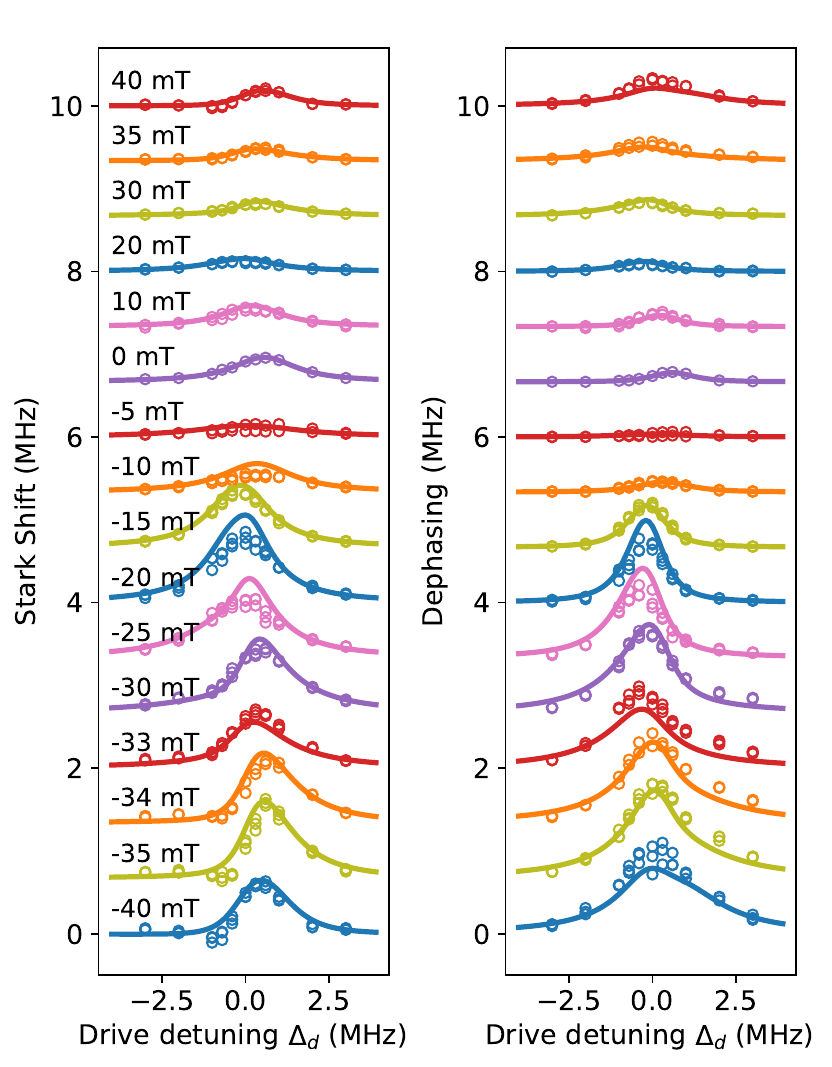}
    \caption{\textbf{Complete data set for parameter-free verification of the master equation model with continuous cavity drive.} The theory prediction (solid lines) and experimental result (dots) of the qubit Ramsey measurement under a continuous cavity drive, as shown in main text Fig.~\ref{fig:cw}.
    The left panel depicts the qubit Stark shift and right panel the dephasing rate under different fields. Note that the theory prediction contains no free parameters and shows good agreement to the experimental data.  
    }
    \label{fig:cw_all}
\end{figure}

\subsection{Validation of the master equation predictions with continuous wave (CW) cavity drive}

In Fig.~\ref{fig:cw_all}, we show the complete data set for qubit Ramsey measurements under a continuous cavity drive, as shown in main text Fig.~\ref{fig:cw}. We include the theory prediction and experimental results for all of the fields, to highlight their agreement and thus provide more support for the parameter-free verification of the master equation model. 

\begin{figure}[tbp]
    \centering
    \includegraphics[width=8.5cm]{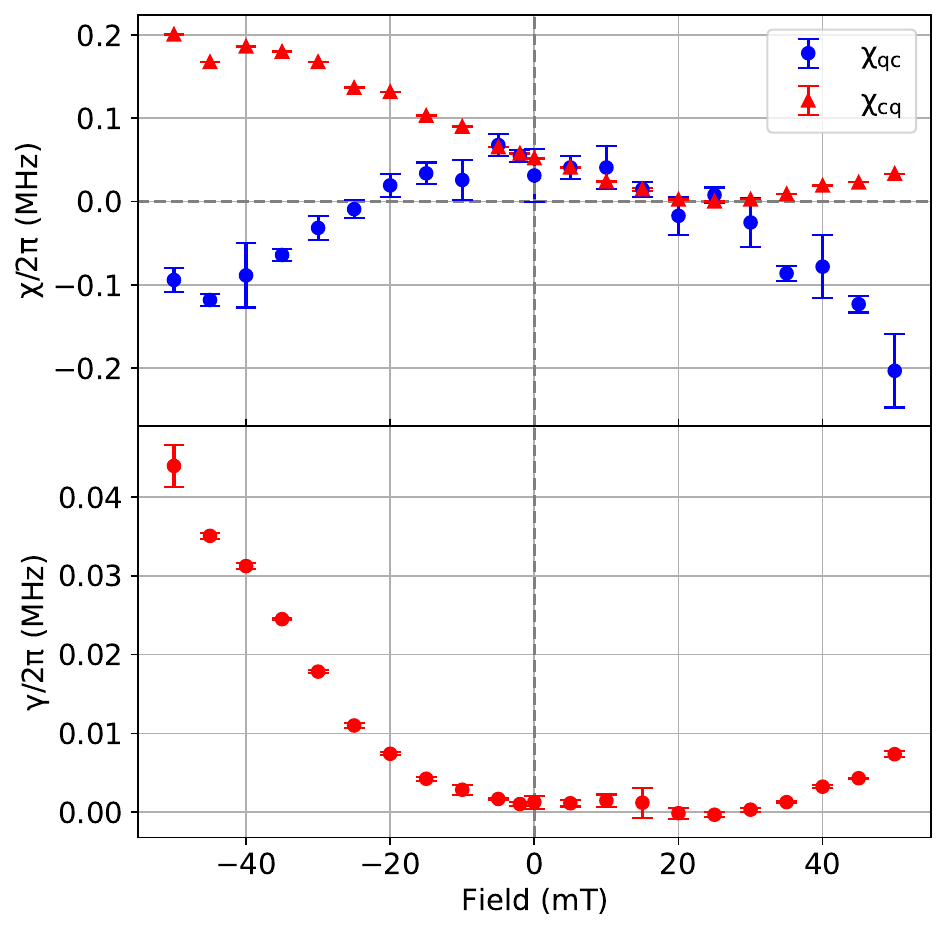 }
    \caption{\textbf{Non-reciprocal frequency shift and dephasing result of setup 2.} The upper panel plots measured $\chi$s at both directions under different external magnetic fields, showing variable non-reciprocity between the cavity and the qubit. The $\chi_{qc}$ is near-symmetric across the external field and the $\chi_{cq}$ is asymmetric. The lower panel is the qubit dephasing rate under different external magnetic fields.
    }
    \label{fig:chis_extra}
\end{figure}
\section{Experimental results on Device B}


\begin{figure}[tbp]
    \centering
    \includegraphics[width=8.5cm]{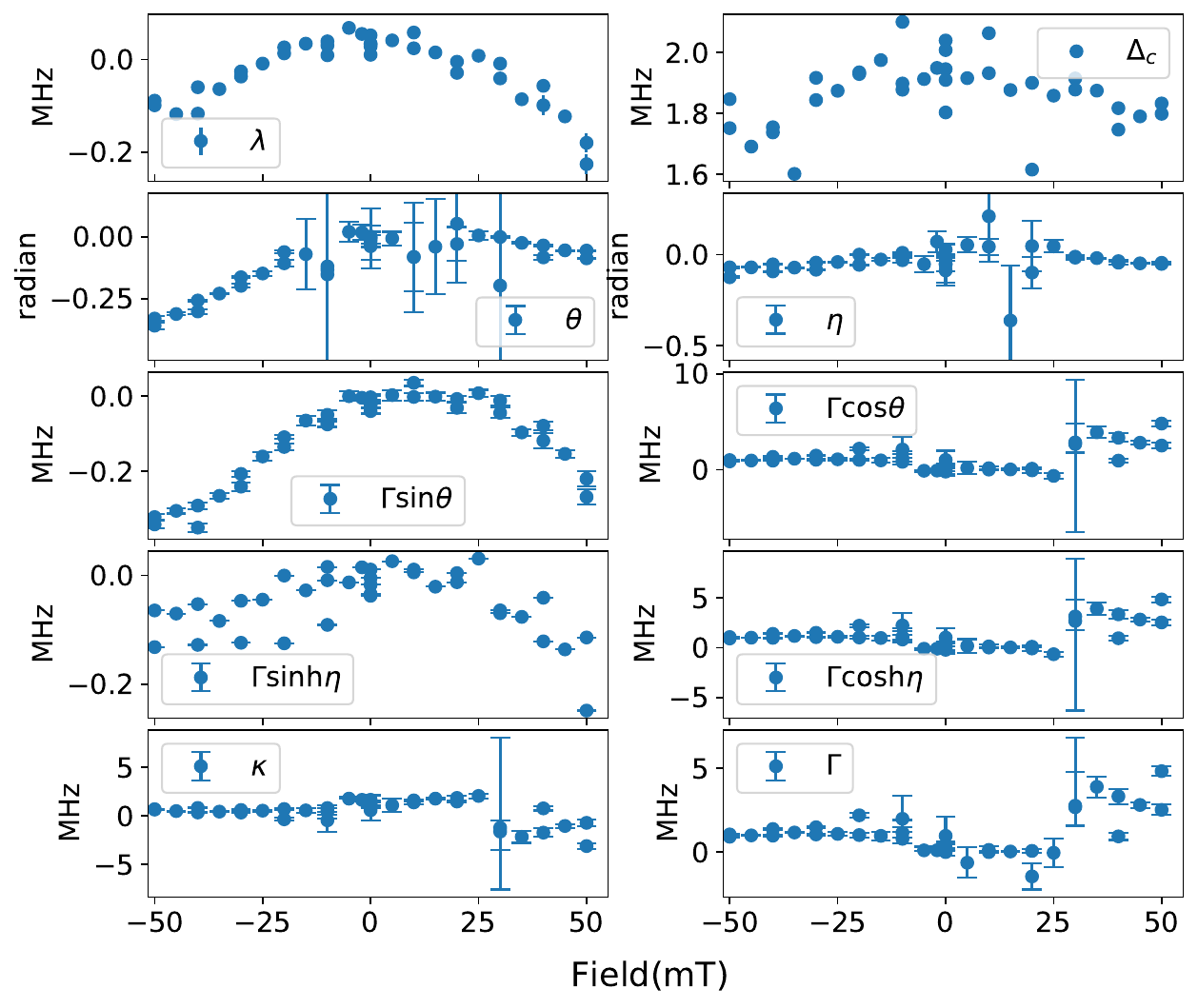}
    \caption{\textbf{Master equation parameters extracted for Device B.} The extracted master equation parameter values and uncertainties for setup 2 utilizing the same Monte Carlo simulation method as the first set of experiments.}
    \label{fig:params_old}
\end{figure}

\begin{figure}[tbp]
    \centering
    \includegraphics[width=8.5cm]{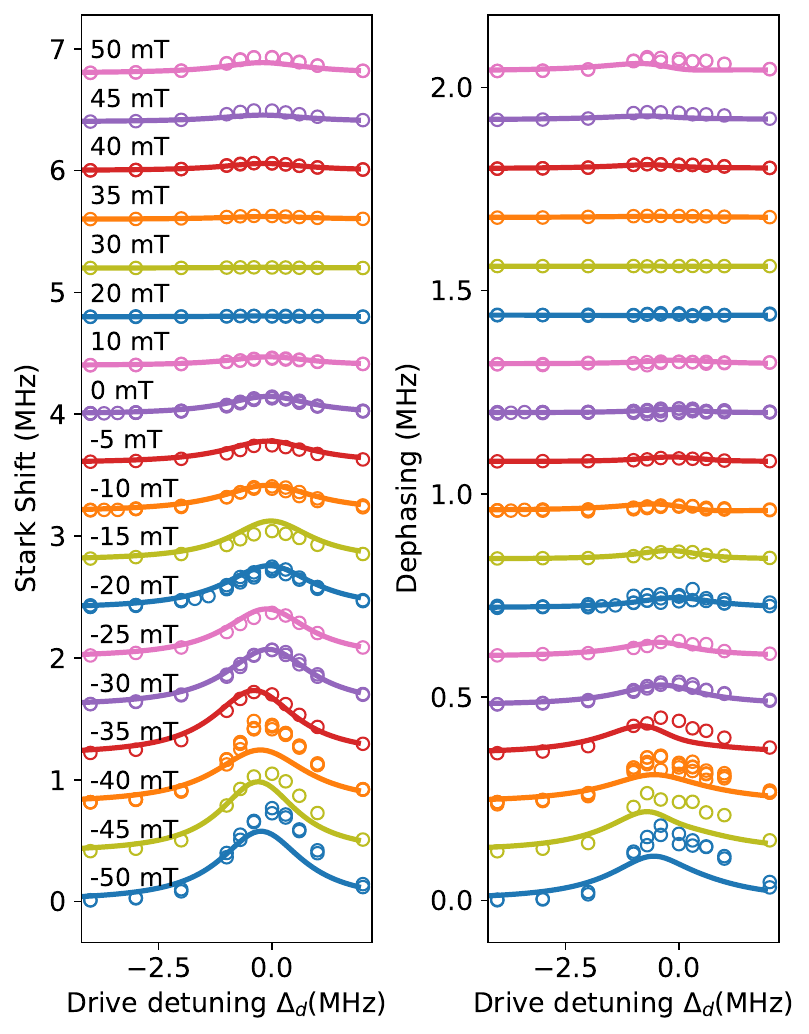 }
    \caption{\textbf{Complete data set of setup 2 for parameter-free verification of the master equation model with continuous cavity drive.} The theory predictions (solid lines) and experimental results (dots) of the setup 2 qubit Ramsey measurement under a continuous cavity drive (see main text Sec.~\ref{main_sec:verif.drives} for discussion on experimental protocol and analysis used). 
    Note that the theory predictions contain no free parameters and shows good agreement with the experimental data for field traces at lower fields, i.e.~within 30 mT.  
    }
    \label{fig:cw_all_extra}
\end{figure}


In addition to the results shown in the main text, we performed a set of measurements on a slightly different device, which we will refer to as Device B. This second setup has similar cavity parameters and different qubit parameters, the main difference being the much smaller dispersive shift $\chi$ between the qubit we are interested in and the broadband cavity that is housing it. The whole set of qubit and cavity parameters for both setups are shown in Table~\ref{table:params}. 

\begin{table}[h!]
\centering
\begin{tabular}{c c c c } 
\hline
&properties &  Device A  &  Device B  \\
\hline
\hline
\multirow{4}{*}{Qubit}&
frequency  & 9.141 GHz & 8.305 GHz \\ 
&$T_1$ & 5.3 $\mu$s& 12.7 $\mu$s\\ 
&$T_2$ & 2.2 $\mu$s& 5.5 $\mu$s\\ 
&$T_{2e}$ & 2.7 $\mu$s& 10.1 $\mu$s\\ 
\hline
\multirow{3}{*}{buffer cavity}&
frequency & 10.808 GHz& 10.812 GHz\\
&linewidth & $>$ 5 MHz& $>$ 5 MHz\\
&$\chi$ & 5.0 MHz& 0.7 MHz\\
\hline
\hline
\multirow{2}{*}{Cavity}&
frequency  & 10.809 GHz & 10.814 GHz \\ 
&linewidth  & 1.7 MHz & 1.8 MHz\\
\hline
\multirow{5}{*}{ancilla qubit}&
frequency  & 8.277 GHz & 9.173GHz \\ 
&$T_1$ & 10.2 $\mu$s& 0.6 $\mu$s\\ 
&$T_2$ & 2.9 $\mu$s& 0.8 $\mu$s\\ 
&$T_{2e}$ & 3.7 $\mu$s& 0.8 $\mu$s\\ 
&$\chi$ & 1.1 MHz& 1.1 MHz\\[1ex] 
\hline
\end{tabular}
\caption{Experimental parameters for the two devices used in the measurements.}
\label{table:params}
\end{table}

We observe the non-reciprocal frequency shift from setup 2, as shown in Fig.~\ref{fig:chis_extra}. We then repeat the procedure of master equation parameter extraction on this setup,
and plot the results in Fig.~\ref{fig:params_old}, to provide further verification for the master equation~\eqref{eq:master_eq}.
There is a discrepancy from the prediction that $\kappa$ is symmetric based on theory as it appears to have a degree of asymmetry in our measurement results. We believe this could be explained by the higher frequency modes that exist in the cavity and can be excited by the cavity drive as well. These modes are coupled to the complex intermediary modes that dephase and Stark shift the qubit. We believe those effects induced by the high frequency modes are symmetric with respect to different external fields. At negative fields, the dephasing and Stark shift effects from our aimed cavity is large, making this extra contribution negligible. However, at positive fields, such contribution from the higher frequency modes could become dominant. This would result in an overestimation of $\Gamma \cosh \eta$ that depends on qubit dephasing measurements, and an over- or underestimation of the absolute value of the non-reciprocal frequency shift $\Gamma \sin \theta$, which depends on the difference of $\chi_\mathrm{qc}$ and $\chi_\mathrm{cq}$. The net effect amounts to a mischaracterization of $\Gamma$ at negative external fields, which in turn leads to a discrepancy in the extracted value of $\kappa$ (see Eq.~\eqref{eq:kappa_eff}). We believe both the deviation of measured $\kappa$ (in both setups) from the symmetric field dependence as predicted by theory, as well as the unphysical values of $\kappa$ extracted for setup 2, could be attributed from the aforementioned effect.

With the fully extracted model, we theoretically compute the qubit Stark shift and dephasing rate under continuous cavity drive and compare them to experimental results.
Interestingly, the master equation still produces relatively reasonable predictions, even though $\kappa$ can now take unphysical values.
Moreover, as Fig.~\ref{fig:cw_all_extra} shows, 
the theoretical lineshapes of qubit frequency shift and dephasing exhibit a nontrivial asymmetry with respect to drive detunings, which matches the asymmetry observed in experimental results. 
This is reasonable as what controls the cavity decay is the sum, ${ \kappa + \Gamma  e ^{  \eta \sigma_z  }}$, which depends on qubit state but in principle does not require precise knowledge of $\kappa$ on its own.



\end{document}